\documentclass[a4paper]{spie}  
\usepackage{graphicx, subfigure}
\usepackage{float}
\usepackage{epsfig}
\usepackage{gensymb}
\title{Analytical computation of stray light in nested mirror modules for X-ray telescopes} 

\author{Daniele~Spiga
\skiplinehalf
INAF/Brera Astronomical Observatory, Via Bianchi 46, 23807 Merate, Italy}
\authorinfo{e-mail: \underline{daniele.spiga@brera.inaf.it}, phone: +39-039-72320427}   
  
\begin{document} 
\maketitle 

\begin{abstract}
Stray light in X-ray telescopes is a well-known issue. Unlike rays focused via a double reflection by usual grazing-incidence geometries such as the Wolter-I, stray rays coming from off-axis sources are reflected only once by either the parabolic or the hyperbolic segment. Although not focused, stray light may represent a major source of background and ghost images especially when observing a field of faint sources in the vicinities of another, more intense, just outside the field of view of the telescope. The stray light problem is faced by mounting a pre-collimator in front of the mirror module, in order to shade a part of the reflective surfaces that may give rise to singly-reflected rays. Studying the expected stray light impact, and consequently designing a pre-collimator, is a typical ray-tracing problem, usually time and computation consuming, especially if we consider that rays propagate throughout a densely nested structure. This in turn requires one to pay attention to all the possible obstructions, increasing the complexity of the simulation. In contrast, approaching the problems of stray light calculation from an analytical viewpoint largely simplifies the problem, and may also ease the task of designing an effective pre-collimator. In this work we expose an analytical formalism that can be used to compute the stray light in a nested optical module in a fast and effective way, accounting for obstruction effects.
\end{abstract}


\section{Introduction}\label{sec:intro} 
\begin{figure}[hbt]
	\centering
	\includegraphics[width = 0.85\textwidth]{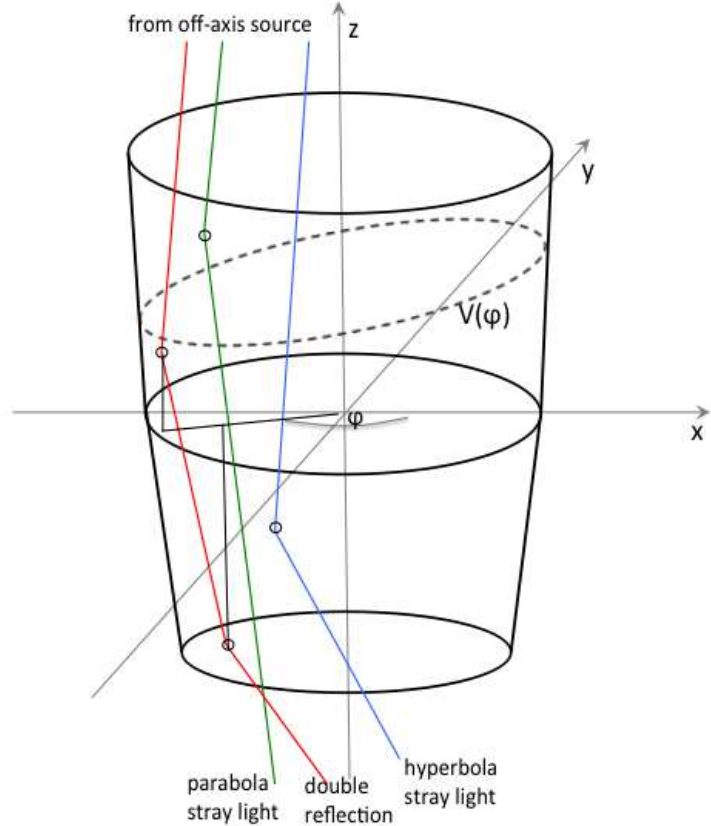}
	\caption{Origin of stray light in an unobstructed Wolter-I mirror shell. Off-axis rays reflected by the primary segment above the $V(\varphi)$ line (Eq.~\ref{eq:V_coef}) do not make the double reflection. Stray light also stems from a direct reflection below the $z$~=0 plane.}
	\label{fig:straylight}
\end{figure}

Optical modules for X-ray telescopes are usually double reflection systems, like the widespread Wolter-I design\cite{VanSpey}. Reflecting X-rays twice halves the focal length and largely suppresses the coma aberration, enabling more compact spacecrafts and larger field of views, at the sole cost of some reduction of the effective area with respect to a single reflection. When a Wolter-I mirror module is illuminated by an on-axis source at infinite distance, all the rays that are reflected by the parabolic segment also impinge onto the hyperbolic segment. But things may change when the source is off-axis or at finite distance: some X-rays can make a single reflection on the parabola, while others can directly impinge on the hyperbola (Fig.~\ref{fig:straylight}). In both cases, they can reach the focal plane without being focused and increase the background or generate "ghost" images. These rays, usually referred to as "stray light", are a well-known problem in X-ray astronomy as they can seriously hamper the observation of faint targets by contamination from intense X-ray objects just {\it outside} the telescope field of view, like e.g., the Crab nebula or even the Sun if the detectors are not shielded against the visible or the infrared light\cite{Peterson1997}. 

When designing X-ray optical modules, the assessment of the stray light contamination is an important step in order to study possible countermeasures. Mirror modules consist of densely nested mirror shells, hence part of stray rays is blocked by the rear (usually non-reflective) side of the inner shells. However, the mirror nesting cannot be too tight, or also doubly-reflected rays from off-axis sources will be obstructed, at the expense of the effective area for imaged sources within the field of view. For this reason, other solutions have been devised out, like X-ray precollimators (see, e.g.\cite{Cusumano2007, Mori2012}) to prevent X-rays from reaching the optical surfaces from directions that may generate stray rays. These auxiliary items are carefully designed, manufactured and aligned to the mirror aperture\cite{eROSITAbaffle}, but at the same time to minimize the obstruction of the effective area for double reflection. Because of the complexity of the possible paths followed by X-rays (stray or not) throughout an optical module, the design and the performance verification of an X-ray optical module and of the pre-collimator is usually done via ray-tracing. This task can be computationally intense and time consuming, especially because the design performance has to checked until an optimal solution is found. In contrast, approaching the same problem from an analytical viewpoint would be useful. Not only to compute in an easy and fast way the stray light impact on a given X-ray module design, but also to find the optimal configuration for the mirror module and the precollimator without the need of writing complex ray-tracing routines. 
 
In previous papers\cite{Spiga2009, Spiga2010} we had already faced the problem of off-axis effective area computation for mirror shells, using analytical expressions. The results could be extended to the case of shells nested in mirror modules, accounting for the mutual obstruction\cite{Spiga2011}. The effective area, as a function of the off-axis angle and the X-ray wavelength, can be expressed by integral equations, with results in excellent agreement with the findings of ray-tracing. In addition, the numerical integration of the analytical formulae requires a time that is orders of magnitudes lower than ray-tracing, and without being affected by statistical uncertainties. In this paper we show that the same formalism can be easily extended to the computation of the {\it effective area for stray light} for an off-axis source.

In Sect.~\ref{sec:recall} we briefly recall the analytical theory of the effective area\cite{Spiga2009, Spiga2011}. This will introduce us to some concepts that we can use to write analytical expressions of the effective area for doubly-reflected rays and, in Sect.~\ref{sec:straylight}, of the effective area for stray light. The expressions for the stray light off the primary and the secondary segment are different and, for brevity, we refer to the former as "primary stray light" and to the latter as "secondary stray light". In Sect.~\ref{sec:valid} we show examples of computation, validating the results by comparison with the findings of a ray-tracing routine. In Sect.~\ref{sec:geometric} we solve the integral equations for the ideal case of a mirror with 100\% reflectivity, obtaining algebraic expressions for the stray light geometric area, in a completely similar way as we did for the double-reflection area\cite{Spiga2009, Spiga2011}. Results are briefly summarized in Sect.~\ref{sec:conclusions}.

We explicitly remark that the formalism for the focused effective area and for stray light are both developed in {\it double cone approximation}. As discussed in detail\cite{Spiga2009}, we are allowed to do this owing to the shallow incidence angles. While the longitudinal curvature of the Wolter-I profile is crucial to concentrate X-ray to a focus, this affects the effective area only to a small extent. For example, in a double cone geometry the incidence angle for a infinitely distant source on-axis is a constant, $\alpha_0$, throughout the entire surface of the primary and the secondary segments. In a real Wolter-I profile, they exhibit a small variation $\Delta \alpha$, related to the curvature of the profile. Fortunately, in grazing incidence geometry it is possible to prove that $\Delta \alpha/\alpha_0 \approx L/4f$, where $L$ is the length of the single segment of the mirror and $f$ its focal length. In practice, the error we make assuming a constant incidence angle is below a few percent in real cases. It can also be proven\cite{Spiga2009} that also the estimation of factors affecting the effective area (e.g., the vignetting) are still on the order of the $L/f$ ratio; therefore, the double cone approximation can be safely applied in the effective area computation. In this work we assume that the double cone approximation can be applied to the stray light theory within a relative error of $L/2f$, which in worst cases amounts to a few percent.

Finally, for simplicity we assume the shells to be continuous at the intersection plane. The theory is easy to extend to the case of primary and secondary segments separated by a gap, but we are not reporting it here, in order to avoid a complication of the expressions.

\section{The off-axis effective area of an X-ray mirror shell} \label{sec:recall}
We now consider a pair of nested, integral mirror shells, with optical axis parallel to the $z$-axis (see Fig.~\ref{fig:obst_scheme}, A). We limit ourselves to the case of mirror shells having all the same primary/secondary intersection plane, and we define it to be the $xy$ plane (therefore, the theory is not directly applicable to a Wolter-Schwarzschild configuration). 

When the shells are illuminated by a distant X-ray source, the inner shell may cause an {\it obstruction} of rays focused by the outer one, and so reduce its effective area. We therefore refer to the outer shell as "reflective" and to the inner shell as "obstructing". Doing this, we implicitly assume that the reflective shell can be uniquely obstructed by the next shell with smaller diameter in the mirror module. For generality, however, we assume the primary and the secondary segments to have different lengths along the optical axis. We denote these lengths as $L_1$, $L_2$ for the reflective shell, and $L_1^*$, $L_2^*$ for the obstructing shell. If $L_1 = L_2$ or $L_1^* = L_2^*$, then we denote their common value with $L$ and $L^*$, respectively. The radii at the primary-secondary intersection plane are $R_0$ for the {\it inner} side of the reflective shell and $R_0^*$ for the {\it outer} side of the obstructing shell. The respective radii at the entrance pupil are denoted as $R_{\mathrm M}$ and $R_{\mathrm M}^*$, while the radii at the exit pupil are $R_{\mathrm m}$ and $R_{\mathrm m}^*$. Finally, be $\alpha_0$  and $\alpha_0^*$ the respective on-axis incidence angles on the primary mirror, at the intersection plane (Fig.~\ref{fig:obst_scheme}, B). Should rays impinge from the source on the secondary segment (and this is really unwanted since it is one of the possible sources of stray light), the incidence angles are $3\alpha_0$  and $3\alpha_0^*$. We assume the source to be in the $xz$ plane, on the side of positive $x$'s, at $z = D$ (either positive or negative, but always $|D| \gg L$ and $|D| \gg f$). Finally, we denote with $\delta = R_0/D$ the beam half-divergence seen by the reflective shell. 

In this section we recall the expressions for the double-reflection effective area of the {\it reflective} mirror shell, seen by the source off-axis by a small angle $\theta > 0$, at the X-ray wavelength $\lambda$, and accounting for the obscuration by the {\it obstructing} shell\cite{Spiga2011}. We denote it with $A_D(\lambda, \theta)$, with the special case of a source at infinity, $A_{\infty}(\lambda, \theta)$. When we refer to the geometric area (i.e., assuming a coating reflectivity $r_{\lambda}(\alpha) =1$), we simply write $A_D(\theta)$ (or $A_{\infty}(\theta)$ for the astronomical case). 

We will see in Sect.~\ref{sec:straylight} that the theory can be easily extended to the computation of a similar effective area for the stray light, i.e., for singly-reflected rays from an off-axis source. This in turn allows one to compute easily the amount of stray light contamination in the field of view of the X-ray telescope if the intensity of the source is known. We follow the aforementioned notation to denote the effective area for stray light, only adding the "SL" superscript.

\subsection{Vignetting coefficients} \label{sec:vignet}
We henceforth recall some useful quantities that enter the computations of the mirror shell effective area\cite{Spiga2009, Spiga2011}. The key concept is the {\it vignetting coefficient} $V$, meant as the fraction of a mirror length left clear by a given factor of obstruction. When we refer to the blocked length fraction for the same reason, we denote it as {\it obstruction coefficient} (i.e., $1-V$). 

\begin{figure}[hbt]
	\centering
	\begin{tabular}{ll}
		\includegraphics[width = 0.46 \textwidth]{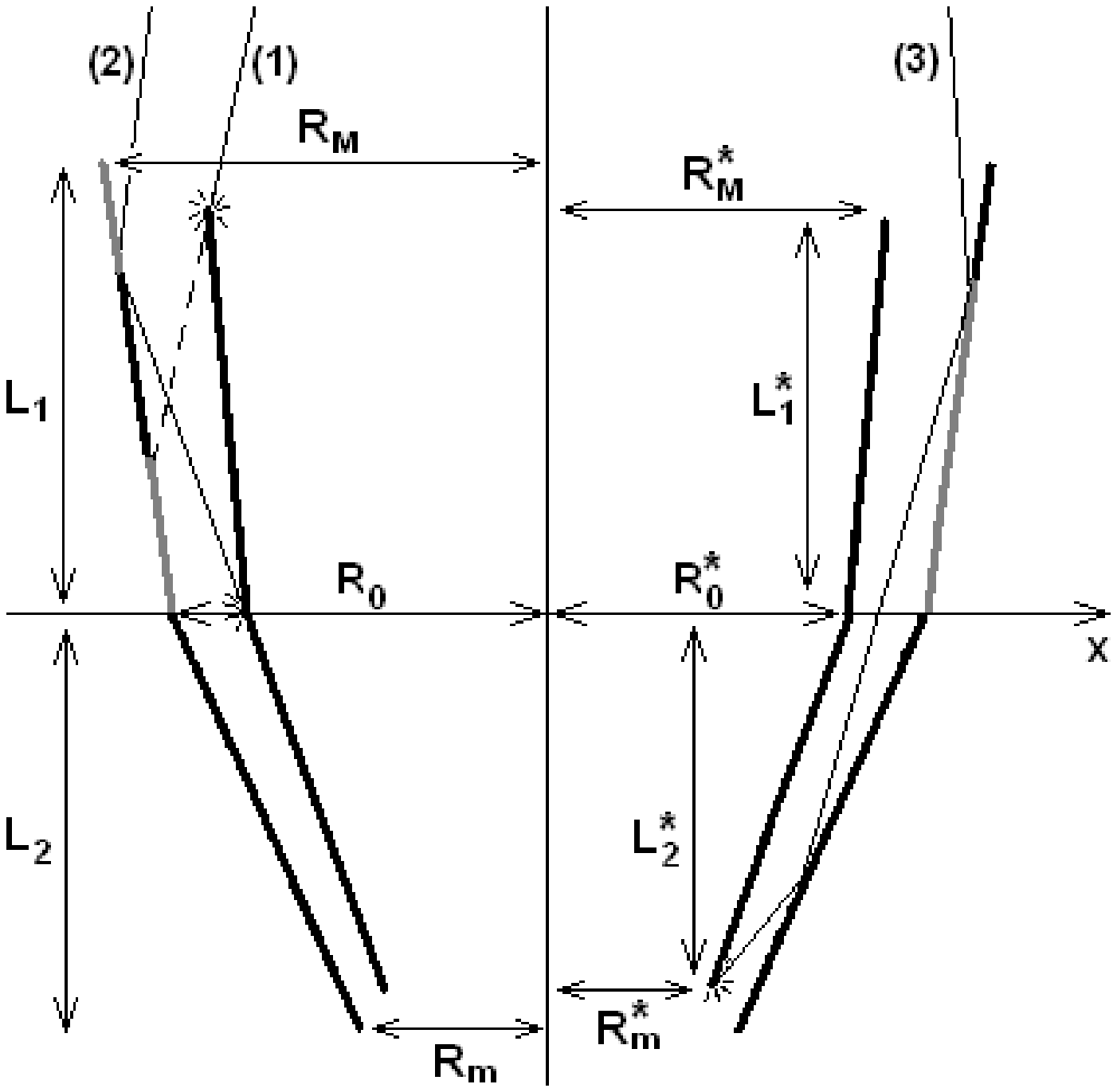}&
		 \includegraphics[width = 0.46 \textwidth]{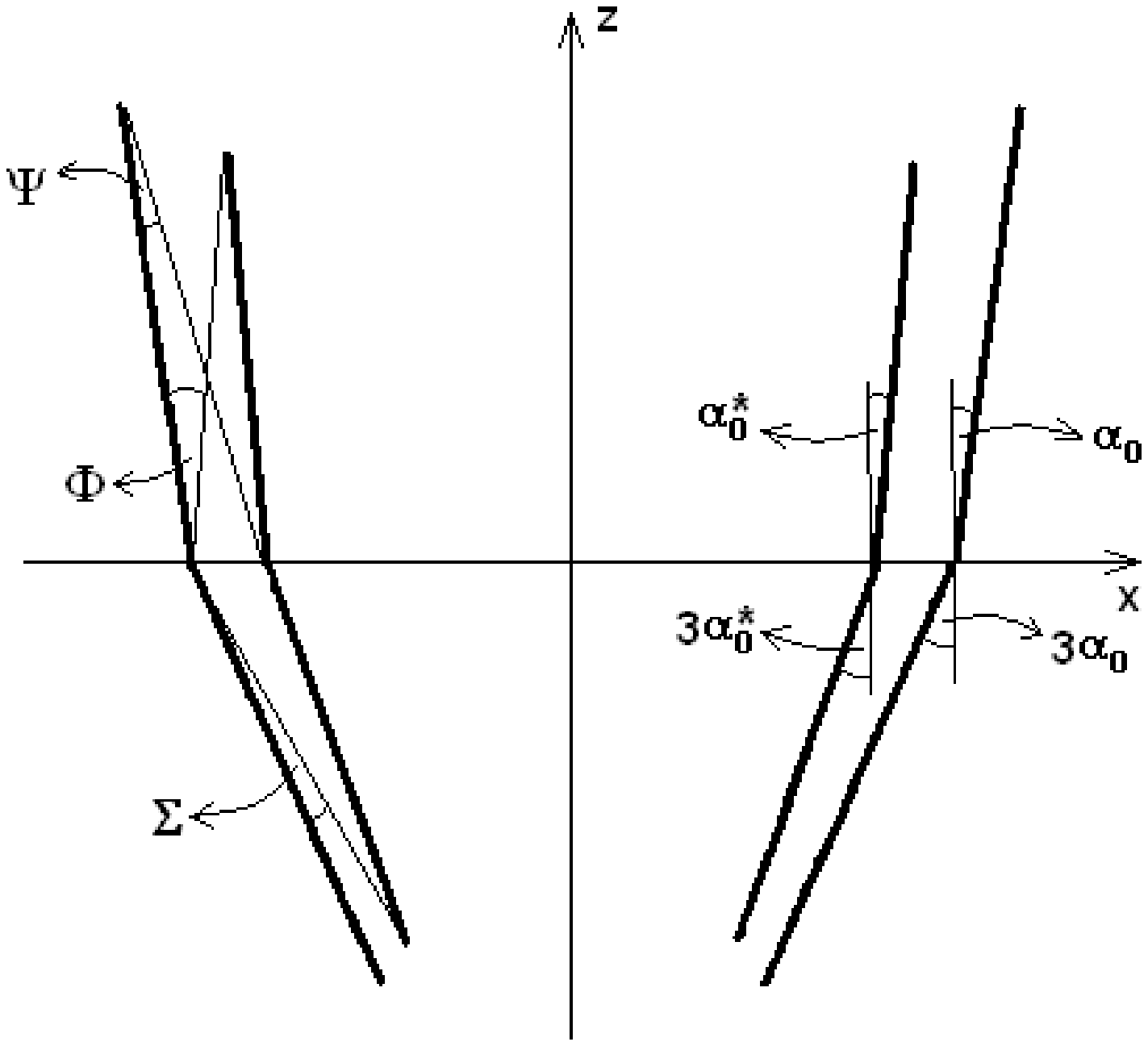}\\
		 \footnotesize A) &  \footnotesize B)
	\end{tabular}
	\caption{A) three possible causes for obstruction of {\it doubly-reflected rays} in a Wolter-I mirror shell: (1) before the first reflection, at the entrance pupil: (2) before the second reflection, on the rear side of the primary segment of the obstructing shell: (3) after the second reflection, on the rear side of the secondary segment of the obstructing shell. The shaded regions are grayed.  B) geometrical meaning of the obstruction parameters (after\cite{Spiga2011}).}
	\label{fig:obst_scheme}
\end{figure}

\begin{enumerate}

\item{A focused ray undergoes two reflections in sequence: the first one on the primary mirror at the grazing angle $\alpha_1$ and the second one on the secondary at the grazing angle $\alpha_2$. In the assumed double-cone approximation, if the source is on-axis we have $\alpha_1 = \alpha_2 = \alpha_0$ throughout the entire mirror surface. If the source is off-axis by $\theta$, the incidence angles vary with the polar angle $\varphi$, measured in the $xy$ plane from the $x$ axis\footnote{More exactly, we should be speaking of "polar angles" because the reflections on the primary and the secondary occur at slightly different values $\varphi_1$ and $\varphi_2$. However, it can be demonstrated\cite{Spiga2009} that this difference is negligible for our scopes, so we simply denote the nearly-common value of the two polar angles with $\varphi$.}:
\begin{eqnarray}
	\alpha_1(\varphi) = \alpha_0+\delta-\theta\cos\varphi, \label{eq:a1}\\
	\alpha_2(\varphi) = \alpha_0-\delta+\theta\cos\varphi. \label{eq:a2}
\end{eqnarray}
We notice that $\alpha_1 +\alpha_2 = 2\alpha_0$ and that the behavior of $\alpha_1$ and $\alpha_2$ is symmetric with respect to the $x$-axis. The expressions are valid as far as they are non-negative.}

\item{We define $V$, the {\it double reflection vignetting coefficient} at the $\varphi$ polar angle as the fraction of primary mirror length from which reflected rays impinge onto the secondary segment. A detailed analysis\cite{Spiga2009} shows that the $V$ coefficient has the expression
\begin{equation}
	V(\varphi) = \frac{L_2\alpha_2}{L_1\alpha_1}.
	\label{eq:V_coef}
\end{equation}
The part of primary mirror that contributes to the double reflection is located on the side of the intersection plane (Fig.~\ref{fig:straylight}). Equation~\ref{eq:V_coef} is valid for non-negative $\alpha_1$ and $\alpha_2$, otherwise it should be set to zero. Moreover, $V=1$ if $L_2\alpha_2 > L_1\alpha_1$.}

\item{The focused beam can be obstructed in three different ways, shown in Fig.~\ref{fig:obst_scheme}, A. Obstruction occurs when the incidence angles exceed three angles that characterize the spacing left between the reflective and the obstructing shell, the {\it vignetting parameters} (Fig.~\ref{fig:obst_scheme}, B):
\begin{equation}
	\Phi = \frac{R_0-R_{\mathrm M}^*}{L_1^*}+\alpha_0,\hspace{2cm}\Psi =\frac{R_0-R_0^*}{L_1}, \hspace{2cm} \Sigma = \frac{R_0-R_{\mathrm m}^*}{L_2^*}-3\alpha_0.
	\label{eq:vig_par}
\end{equation}
If $L_1^* = L_1 = L_2^*$ and $f \gg L_1$, then $\Phi \approx \Psi \approx \Sigma$.}

\item{The first kind of obstruction occurs when rays are blocked before they can make the first reflection: the primary segment of the obstructing shell casts a shadow on the reflective shell near the intersection plane (Fig.~\ref{fig:obst_scheme}, A). The residual fraction of illuminated primary mirror at $\varphi$ is located on the side of the entrance pupil and expressed by the $V_1$ coefficient:
\begin{equation}
	V_1(\varphi) = 1+\frac{L_1^*(\Phi- \alpha_1)}{L_1\alpha_1}.
	\label{eq:V1_coef}
\end{equation}
The obstruction of this species is always maximum at $\varphi \approx \pi$, where $\alpha_1$ is larger.}

\item{The second kind of obstruction may occur after the first reflection, if rays are blocked by the rear side of the obstructing shell at $z > 0$. Just like $V$, the fraction of primary segment that is left clear is on the side of the intersection plane and expressed by the $V_2$ coefficient,
\begin{equation}
	V_2(\varphi) = \frac{\Psi}{\alpha_1}.
	\label{eq:V2_coef}
\end{equation}
Also in this case, the obstruction becomes more severe for large values of $\alpha_1$. Fortunately, mirror modules can be designed to be completely unaffected by this kind of obstruction\cite{Spiga2011} at any value of $\theta$.}

\item{The third kind of obstruction may occur after the second reflection, when the ray impacts on the rear side of the secondary segment of the obstructing shell. In this case, the corresponding fraction of primary mirror $V_3$ that is not obscured is on the side of the entrance pupil (just like $V_1$) and provided by the expression
\begin{equation}
	V_3(\varphi) = 1+\frac{L_2^*(\Sigma- \alpha_2)}{L_1\alpha_1}.
	\label{eq:V3_coef}
\end{equation}
This kind of obstruction, unlike the others, becomes important near $\varphi =0$, where $\alpha_1$ is shallower and $\alpha_2$ is larger.}
\end{enumerate}

The explicit dependence of $V$, $V_1$, $V_2$, and $V_3$ on $\varphi$ is obtained substituting the expressions of $\alpha_1$ and $\alpha_2$ (Eqs.~\ref{eq:a1} and~\ref{eq:a2}), always with the constraint that $0 \le V_j \le 1$, and 0 or 1 outside this range. Hence, there is no obstruction of the first kind whenever $\alpha_1 < \Phi$, of the second kind if $\alpha_1 < \Psi$, and of the third kind if $\alpha_2 < \Sigma$. 

\subsection{Expression for the double-reflection effective area} \label{sec:effarea}
Using the vignetting coefficient expressions in Sect.~\ref{sec:vignet}, the general expression for the double-reflection effective area of an obstructed mirror shell can be derived easily (refer to \cite{Spiga2011} for the derivation, with a slight change of notation):
\begin{equation}
	A_D(\lambda, \theta) = 2R_0\int_0^{\pi}[(L\alpha)_{\mathrm{min}} -(L\alpha)_{\mathrm{max}}]_{\ge 0} \,r_{\lambda}(\alpha_1)r_{\lambda}(\alpha_2) \,\mbox{d}\varphi, 
	\label{eq:A_2r_free}
\end{equation} 
where we have set
\begin{eqnarray}
	(L\alpha)_{\mathrm{min}} & = & \min\left(L_1\alpha_1, L_2\alpha_2, L_1\Psi\right),\label{eq:La_min}\\
	(L\alpha)_{\mathrm{max}}  & = &  \max\left[L_1^*(\alpha_1-\Phi), L_2^*(\alpha_2-\Sigma), 0\right]. 	\label{eq:La_max}
\end{eqnarray} 
The focused effective area is therefore {\it completely expressed in terms of the vignetting parameters and the incidence angles}, which in turn are a function of $\varphi$. The brackets $[\,]_{\ge 0}$ mean that the enclosed expression has to be non-negative, otherwise it is set to zero (as it is when either $\alpha_1 < 0$ or $\alpha_2 < 0$). For this reason, the integration of Eq.~\ref{eq:A_2r_free} {\it cannot be carried out separately for the two terms}. 

In the previous formula, $r_{\lambda}(\alpha_1)$, $r_{\lambda}(\alpha_2)$ are the reflectivities of the mirror coating at the X-ray wavelength $\lambda$ on the primary and the secondary segment, respectively. Since the $r_{\lambda}(\alpha)$'s functions are difficult to express in an algebraic form, Eq.~\ref{eq:A_2r_free} is seldom integrated analytically. In contrast, an explicit integration is possible for the geometric area, setting $r_{\lambda}(\alpha)=1$ for any value of $\lambda$ and $\alpha$. Of course, Eq.~\ref{eq:A_2r_free} can {\it always} be integrated numerically.

In Eq.~\ref{eq:A_2r_free}, the integration is extended to the entire range of polar angles, as we reasonably suppose that the focal spot is entirely included in the detector area. Actually, the integration range is reduced to $[0,\pi]$ -- adding a factor of 2 -- because the integrand is symmetric with respect to the $x$-axis. The effective area of the entire module is simply obtained by summing the contributions of pairs of adjacent shells, playing in sequence the role of reflective and obstructing shells. We have hitherto neglected the obstruction of supporting structures ("spiders"), which are, indeed, always present. To preserve the mirror stiffness, the spiders spokes usually have a thickness that increases in proportion with $R_0$, therefore the set of polar angles P they occupy is the same for all the shells in the module. In order to account for the effective area loss, one just has to compute the integral in Eq.~\ref{eq:A_2r} zeroing the integrand at the $\varphi$ values occupied by the spider spokes. More formally, we introduce $\chi_{\mathrm{P}}(\varphi)$, the characteristic function of the P set, and Eq.~\ref{eq:A_2r_free} becomes:
\begin{equation}
	A_D(\lambda, \theta) = 2R_0\int_0^{\pi}[(L\alpha)_{\mathrm{min}} -(L\alpha)_{\mathrm{max}}]_{\ge 0} \,r_{\lambda}(\alpha_1)r_{\lambda}(\alpha_2)\,\chi_{\mathrm{P}} \,\mbox{d}\varphi, 
	\label{eq:A_2r}
\end{equation} 
where the $[\,]_{\ge 0}$ brackets have the same meaning as in Eq.~\ref{eq:A_2r_free}.
\begin{figure}[t]
	\centering
	\includegraphics[width = 0.5 \textwidth]{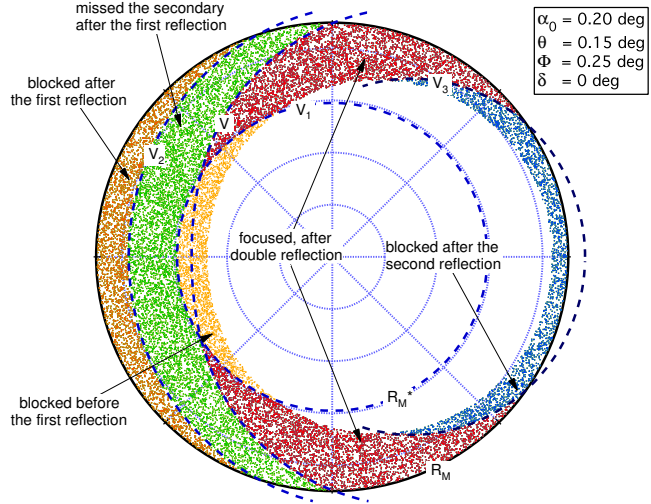}
	\caption{ Initial positions and destinations of 40000 rays at the entrance pupil (points) for a Wolter-I mirror shell with $L = L^*$, and the angular parameters reported in the legend ($\Phi \approx \Psi \approx \Sigma$). The radial scale has been expanded. Only rays that would have struck the primary mirror were traced. The limits of the regions of different vignetting (dashes) are computed from the vignetting coefficients (after\cite{Spiga2011}).}
	\label{fig:entrance_pupil}
\end{figure}

\section{Analytical formalism for stray light}\label{sec:straylight}
\subsection{General expressions}\label{sec:general}
\paragraph{Primary stray light}
From the definition of $V$ (Eq.~\ref{eq:V_coef}), the fraction of primary that does not make the second reflection at the polar angle $\varphi$ is $1-V$. This means that the infinitesimal geometric area for stray light between $\varphi$ and $\varphi+\Delta\varphi$ is $L_1\alpha_1 (1-V) R_0\,\Delta\varphi$. Therefore, in absence of obstructions, the total effective area element for primary stray light {\it would} be $ R_0 L_1\alpha_1 (1-V)r_{\lambda}(\alpha_1)\,\Delta\varphi$ where $r_{\lambda}(\alpha_1)$ is the primary mirror reflectivity.

Indeed, at sufficiently large angles the stray light starts to be blocked at the intersection plane by the rear side of the obstructing shell, i.e., $V_2$ (Eq.~\ref{eq:V2_coef}). A typical situation is depicted in Fig.~\ref{fig:entrance_pupil}: the primary mirror segment region that generates the primary stray light (green dots) is the one on the side of the entrance pupil, mostly in the neighborhoods of $\varphi \approx \pi$. The obstruction of second kind, however, concerns exactly the same region (orange dots) and this is the reason why it is mostly harmless for double reflection. Also the obstruction of the first species (yellow dots) contributes to block the primary stray light, and the relative vignetting is described by $V_1$ (Eq.~\ref{eq:V1_coef}).

\begin{figure}[tbh]
	\centering
	\begin{tabular}{ll}
		\includegraphics[width = 0.48 \textwidth]{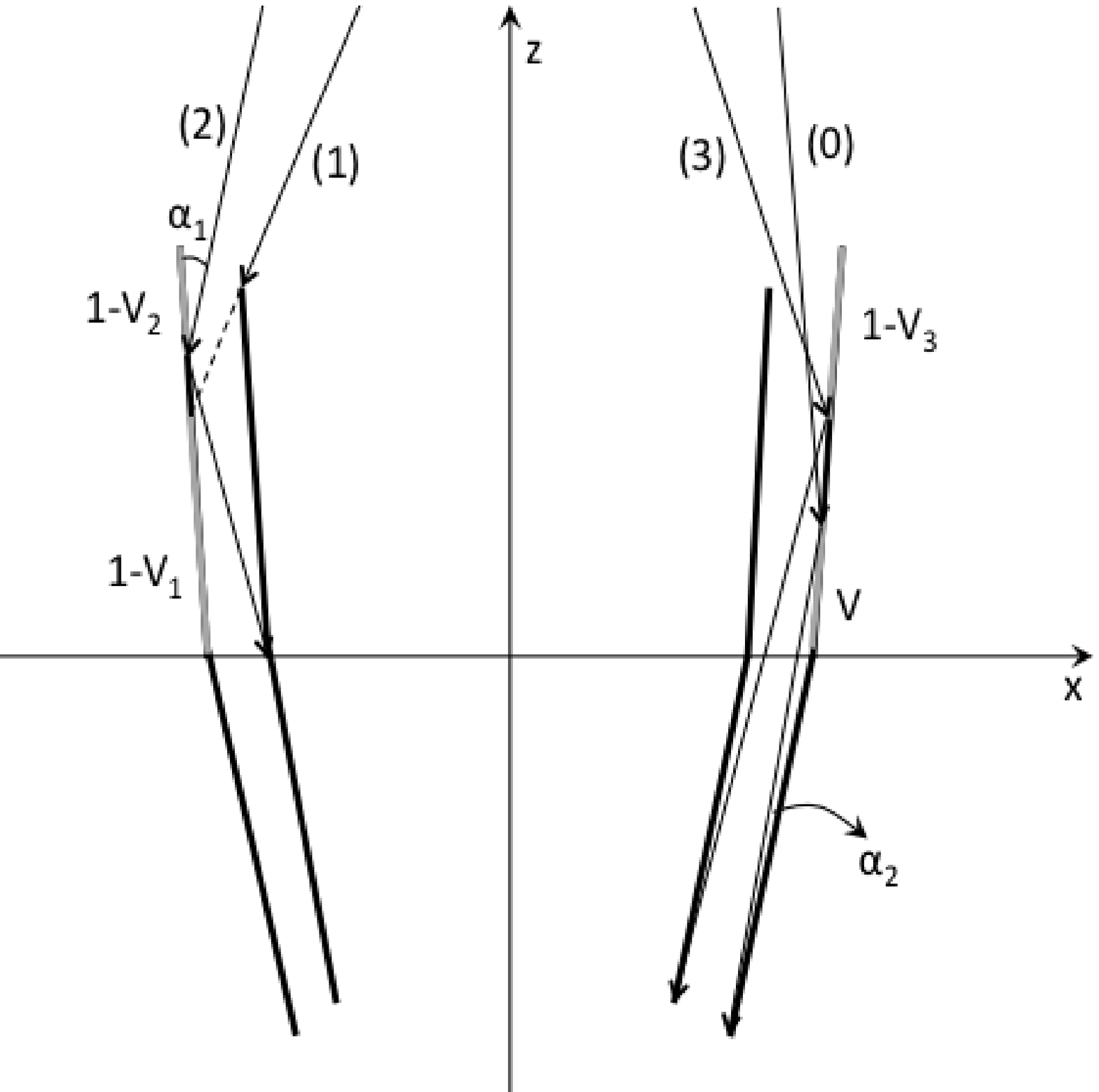}&
		 \includegraphics[width = 0.48 \textwidth]{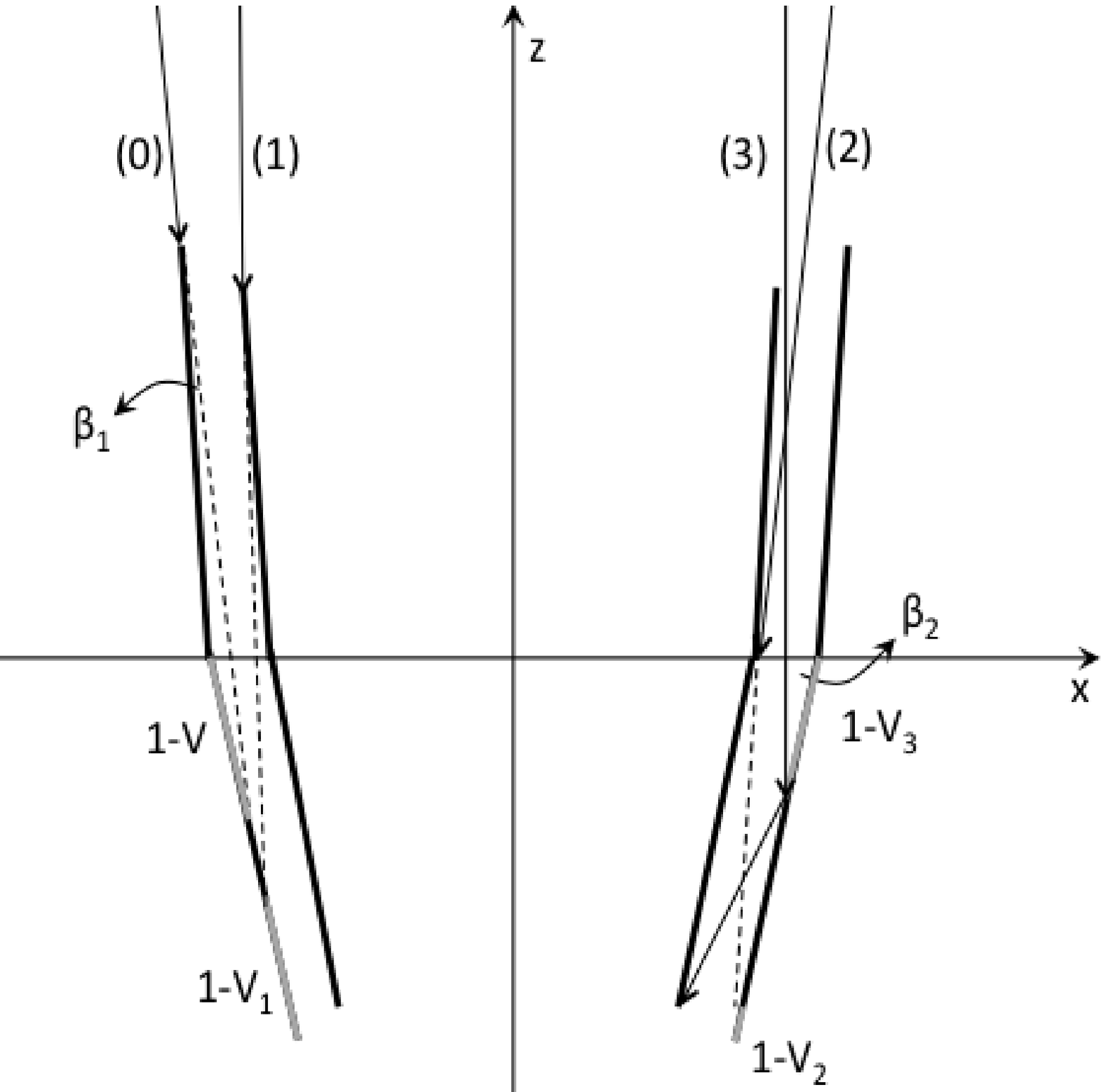}\\
		 \footnotesize A) &  \footnotesize B)
	\end{tabular}
	\caption{A) obstructions of {\it primary stray light} of a Wolter-I mirror shell: (0) for double reflection; (1) at the entrance pupil: (2) on the primary segment of the obstructing shell: (3) on the secondary segment of the obstructing shell.  B) obstructions of {\it secondary stray light}: (0) on the rear side of the primary segment of the reflective shell; (1) by the obstructing shell, on the front side; (2) on the rear side of the obstructing shell, before reflection; (3) on the rear side of the obstructing shell, after reflection. The shaded regions are grayed.}
	\label{fig:SL_obstructions}
\end{figure}

As $V_3$ (Eq.~\ref{eq:V3_coef}) describes the vignetting after the second reflection, it is not directly applicable to this case. Nevertheless, also singly-reflected rays can strike on the rear side of the secondary segment of the blocking shell. This case was not considered in the original paper\cite{Spiga2011} because the obstruction at the exit pupil was therein considered only for doubly-reflected rays. However, the computation can be easily adapted to the primary stray light case. Following passages similar to those in Appendix B.3 in\cite{Spiga2011}, which we omit here, we obtain a modified form of $V_3$:
\begin{equation}
	V_3(\varphi) = \frac{L_2^*(\Sigma+\alpha_2)}{L_1\alpha_1}.
	\label{eq:V3_mod}
\end{equation}
Just like the second one, this kind of obstruction occurs near the entrance pupil (Fig.~\ref{fig:SL_obstructions}, A). 

Summarizing, the non-obstructed length of the primary mirror at $\varphi$ is the least between $V_2$ and $V_3$. From this term, we have to subtract the maximum between the fraction of doubly-reflected rays, $V$, and the obstruction at the intersection plane, $(1-V_1)$, to obtain the total vignetting:
\begin{equation}
	V_{\mathrm{tot},1}= \min(1, V_2, V_3)-\max(V, 1-V_1, 0),
	\label{eq:totV_p}
\end{equation}
where we have added a "1" in the first term and a "0" in the second term to avoid negative obstructions. Replacing now Eqs.~\ref{eq:V_coef}, \ref{eq:V1_coef}, \ref{eq:V2_coef}, and~\ref{eq:V3_mod} into the previous expression, we have
\begin{equation}
	V_{\mathrm{tot},1}= \min\left(1, \frac{\Psi}{\alpha_1}, \frac{L_2^*(\Sigma+ \alpha_2)}{L_1\alpha_1}\right)-\max\left(\frac{L_2\alpha_2}{L_1\alpha_1}, \frac{L_1^*(\alpha_1-\Phi)}{L_1\alpha_1}, 0\right).
	\label{eq:totV_p_bis}
\end{equation}
Since the effective area element between $\varphi$ and $\varphi+\Delta\varphi$ is $V_{\mathrm{tot},1}L_1\alpha_1R_0\Delta\varphi\,r_{\lambda}(\alpha_1)$, we finally obtain the {\it general expression of the effective area for the primary stray light}:
\begin{equation}
	A^{\mathrm{SL,1}}_D(\lambda, \theta) = 2R_0\int_0^{\pi}[(L\alpha)_{\mathrm{min}, 1} -(L\alpha)_{\mathrm{max}, 1}]_{\ge 0} \,r_{\lambda}(\alpha_1) \,\chi_{\mathrm{P}}\,\mbox{d}\varphi, 
	\label{eq:Asl_p}
\end{equation} 
where the superscripts indicate primary stray light, we have included the shading of structures via the characteristic function $\chi_{\mathrm{P}}(\varphi)$ (Sect.~\ref{sec:effarea}), and we have set
\begin{eqnarray}
	(L\alpha)_{\mathrm{min},1} & = & \min[L_1\alpha_1, L_1\Psi, L_2^*(\Sigma+ \alpha_2)],\label{eq:La_min1}\\
	(L\alpha)_{\mathrm{max},1}  & = &  \max[L_2\alpha_2, L_1^*(\alpha_1-\Phi), 0] , 	\label{eq:La_max1}
\end{eqnarray} 
provided that $\alpha_1 \ge 0$. The condition $\alpha_2 \ge 0$ is not necessary because there is no true second reflection, and is anyway fulfilled by the "0" in Eq.~\ref{eq:La_max1}. The $[\,]_{\ge 0}$ brackets have exactly the same meaning as in Eq.~\ref{eq:A_2r}.

Inspection of Eqs.~\ref{eq:La_min1} and~\ref{eq:La_max1} shows that a reduction of the stray light from primary mirror of given length and incidence angle can be achieved by reducing $(L\alpha)_{\mathrm{min}}$, e.g., diminishing $\Psi$ (e.g., via a denser nesting) or $\Sigma$. This can be obtained also by an X-ray baffle located at either the intersection plane or the exit pupil\cite{Cusumano2007}. Another method consists of increasing $(L\alpha)_{\mathrm{max}}$, for example adopting a design with $L_2 > L_1$ that has also the desired effect of increasing the off-axis effective area for focused rays. In contrast, a reduction of $\Phi$ has a minor effect on the primary stray light, since it would chiefly obstruct doubly-reflected rays before blocking the primary stray light (see Fig.~\ref{fig:entrance_pupil}). This is the reason why X-ray baffles at the entrance pupil are not very effective at suppressing primary stray light\cite{Cusumano2007}. 

\paragraph{Secondary stray light}
The treatment of the secondary stray light, i.e., light directly impinging onto the secondary segment, can be derived using exactly the same arguments we used to obtain Eq.~\ref{eq:Asl_p}. The incidence angle, $\beta_2$, clearly differs from Eq.~\ref{eq:a1} as the mirror slope is 3 times as large: 
\begin{equation}
	\beta_2 = 3\alpha_0+\delta-\theta\cos\varphi.
	\label{eq:b1} 
\end{equation}
No second reflection is obviously possible for this species of stray light, but for reasons that will be explained later we also define
\begin{equation}
	\beta_1 = \theta\cos\varphi-\alpha_0-\delta:
	\label{eq:b2} 
\end{equation}
this angle represents the incidence angle on the rear side of the primary segment of the reflective shell. In this case the concept of vignetting for double reflection cannot be applied; hence the usual expression of the $V$ coefficient (Eq.~\ref{eq:V_coef}) is meaningless here. There is, however, the obstruction by the upper edge of the primary segment of the {\it reflective shell}, which starts to occur only when the incidence angle on the primary segment becomes negative and shades the regions near the intersection plane: developing calculations similar to those in the Appendix A of the already cited work\cite{Spiga2009}, we obtain a vignetting coefficient very similar to the usual expression for $V$:
\begin{equation}
	V  =  1-\frac{L_1\beta_1}{L_2\beta_2},
	\label{eq:V_bis}
\end{equation}
and so we keep denoting it with $V$. In addition, there is the usual obscuration by the upper edge of the obstructing shell, which shades the mirror length near the exit pupil ($V_1$). There is still the obstruction on the same side by the mirror kink at the intersection plane ($V_2$), which however has to be modified to account for blocking the direct illumination, not the reflected one. Finally, there is also the obstruction after the reflection, chiefly involving rays reflected near the intersection plane ($V_3$). The vignetting coefficients can be derived using the same method reported in the Appendix B of a previous paper\cite{Spiga2011}, and -- omitting the proofs -- the resulting expressions are
\begin{eqnarray}
	V_1 & = & \frac{L_1^*(\Phi-\alpha_1)}{L_2\beta_2}, \label{eq:V1_bis}\\
	V_2 & = & \frac{L_1\Psi}{L_2\beta_2}, \label{eq:V2_bis}\\
	V_3 & = & 1+\frac{L_2^*(\Sigma-\beta_2)}{L_2\beta_2}. \label{eq:V3_bis}
\end{eqnarray}
We note that in Eq.~\ref{eq:V1_bis} the $\alpha_1$ angle appears instead of $\beta_2$ to account for the direct incidence onto a surface with slope $3\alpha_0$. We now have for the total vignetting
\begin{equation}
	V_{\mathrm{tot},2}= \min(1, V_1, V_2)-\max(1-V, 1-V_3, 0),
	\label{eq:totV_h}
\end{equation}
which becomes, after replacing the corresponding expressions,
\begin{equation}
	V_{\mathrm{tot},2}= \min\left(1, \frac{L_1^*(\Phi-\alpha_1)}{L_2\beta_2}, \frac{L_1\Psi}{L_2\beta_2}\right)-\max\left(\frac{L_1\beta_1}{L_2\beta_2}, \frac{L_2^*(\beta_2-\Sigma)}{L_2\beta_2}, 0\right).
	\label{eq:totV_h_bis}
\end{equation}
Therefore, the {\it general expression of the effective area for the secondary stray light} is
\begin{equation}
	A^{\mathrm{SL,2}}_D(\lambda, \theta) = 2R_0\int_0^{\pi}[(L\beta)_{\mathrm{min},2}-(L\beta)_{\mathrm{max},2}]_{\ge 0} \,r_{\lambda}(\beta_2) \,\chi_{\mathrm{P}} \,\mbox{d}\varphi, 
	\label{eq:Asl_h}
\end{equation} 
where the superscripts, the subscript, and the $\chi_{\mathrm{P}}(\varphi)$ function have the same meaning as in Eq.~\ref{eq:Asl_p}, and where we have set
\begin{eqnarray}
	(L\beta)_{\mathrm{min},2} & = & \min[L_2\beta_2, L_1\Psi, L_1^*(\Phi-\alpha_1)], \label{eq:Lb_min2}\\
	(L\beta)_{\mathrm{max}, 2}  & = &  \max[L_1\beta_1, L_2^*(\beta_2-\Sigma), 0] . 	\label{eq:Lb_max2}
\end{eqnarray} 
provided that, as usual, $\beta_2 \ge 0$. We therefore see that the effective area for the double reflection, for primary stray light, and for the secondary one are provided all by the same formula, only differing by the definition of the terms appearing in there. We also notice that the secondary one can be effectively shaded out by reducing $\Psi$ (denser nesting, baffle at the intersection plane) or $\Phi$ (baffle at the entrance pupil).
\begin{figure}[tbh]
	\centering
		\includegraphics[width = 0.5 \textwidth]{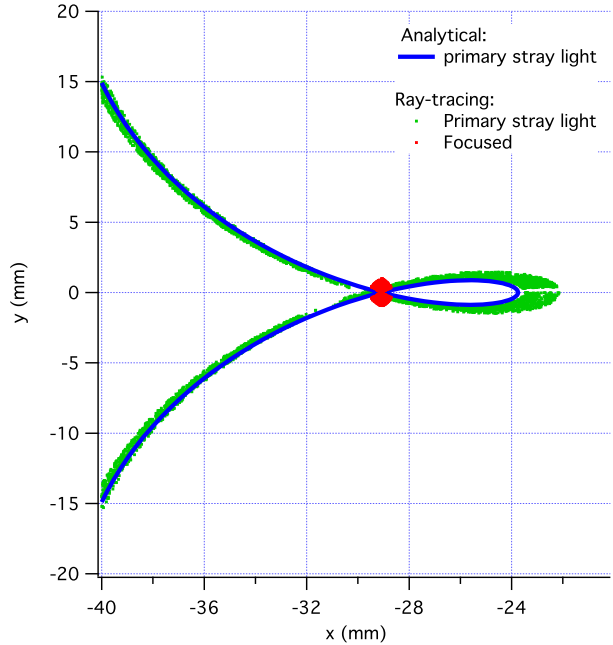}
		\caption{The primary stray light pattern described by a mirror shell with $f$ = 10~m, $R_0$~= 105.5~mm, $\delta= 0$, $\theta$~= 10~arcmin. The analytical curve was computed with Eqs.~\ref{eq:x_polar} and~\ref{eq:y_polar}.}
	\label{fig:SL_pattern}
\end{figure}

\subsection{Stray light within the detector field}\label{sec:inFOV}
As indicated by the integration range $[0, \pi]$, Eqs.~\ref{eq:Asl_p} and~\ref{eq:Asl_h} return the {\it total} stray light, i.e., reflected by the complete mirror shell, albeit limited by the obstruction, and detected over an infinitely extended focal plane. For the focused beam, this is reasonable because the detector can always be supposed to include the complete focal spot. In contrast -- and fortunately -- usually only a part of the stray light from an off-axis source enters the detector, assumed to be a square area of side $d$, located at $z = -f' = -R_0/(4\alpha_0-\delta)$, and centered on the focal spot at $x = -\theta f'$. The effective area within the detector area is the quantity that matters in order to evaluate the effective amount of background generated by an X-ray source at the off-axis angle $\theta$.

The stray light pattern can be described analytically\cite{Spiga2011} using parametric equations as a function of the polar angle $\varphi$:
\begin{eqnarray}
	x_n(\varphi) &=& [R_0-(2(2n-1)\alpha_0+\delta)f']\cos\varphi+\theta f'\cos2\varphi \label{eq:x_polar}\\
	y_n(\varphi) &=& [R_0-(2(2n-1)\alpha_0+\delta)f']\sin\varphi+\theta f'\sin2\varphi \label{eq:y_polar}
\end{eqnarray}
where $n=1$ for the primary (Fig.~\ref{fig:SL_pattern}) and $n=2$ for the secondary stray light respectively. The condition to fulfill is
\begin{equation}
	\min\left[|x_n+\theta f'|, |y_n|\right] \le \frac{d}{2},
	\label{eq:condition}
\end{equation}
and solving Eq.~\ref{eq:condition} for $\varphi$, one obtains the interval of polar angles $[\varphi_{n,-}, \varphi_{n,+}]$ that contribute to the stray light inside the detector, with the constraints $\varphi_{n,-} \ge 0$ and  $\varphi_{n,+} \le \pi$. The effective area for stray light is computed from Eqs.~\ref{eq:Asl_p} and~\ref{eq:Asl_h}, with modified integration limits:
\begin{eqnarray}
	A^{\mathrm{SL,1}}_D(\lambda, \theta) & = &2R_0\int_{\varphi_{1,-}}^{\varphi_{1,+}}[(L\alpha)_{\mathrm{min}, 1} -(L\alpha)_{\mathrm{max}, 1}]_{\ge 0} \,r_{\lambda}(\alpha_1) \,\chi_{\mathrm{P}} \,\mbox{d}\varphi, \label{eq:Asl_p_lim} \\
	A^{\mathrm{SL,2}}_D(\lambda, \theta) & = & 2R_0\int_{\varphi_{2,-}}^{\varphi_{2,+}}[(L\beta)_{\mathrm{min},2}-(L\beta)_{\mathrm{max},2}]_{\ge 0} \,r_{\lambda}(\beta_2)\, \chi_{\mathrm{P}} \,\mbox{d}\varphi, \label{eq:Asl_h_lim}
\end{eqnarray} 
whilst the integrand expressions remain unchanged, and the vignetting by the spider is still included via the $\chi_{\mathrm{P}}(\varphi)$ function.

\section{Examples and result validations via ray-tracing}\label{sec:valid}
In this section we provide some examples of stray light effective area computation. We have implemented Eqs.~\ref{eq:A_2r}, \ref{eq:Asl_p_lim} and~\ref{eq:Asl_h_lim} into an IDL code, and we have checked the correctness of results by comparison with the findings of a detailed ray-tracing routine.

\begin{figure}[htb]
	\centering
	\begin{tabular}{ll}
		\includegraphics[width = 0.48 \textwidth]{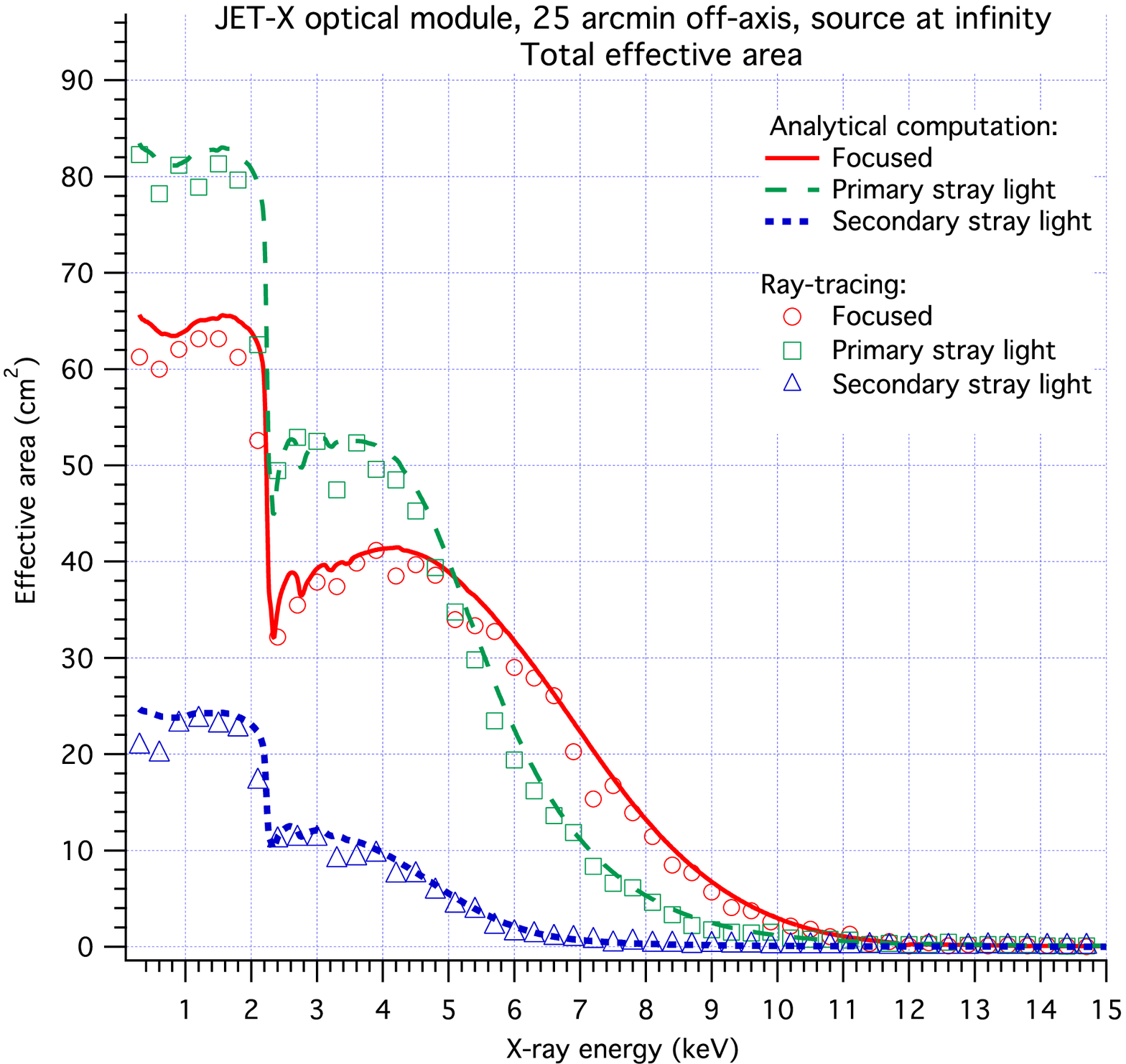}&
		 \includegraphics[width = 0.44 \textwidth]{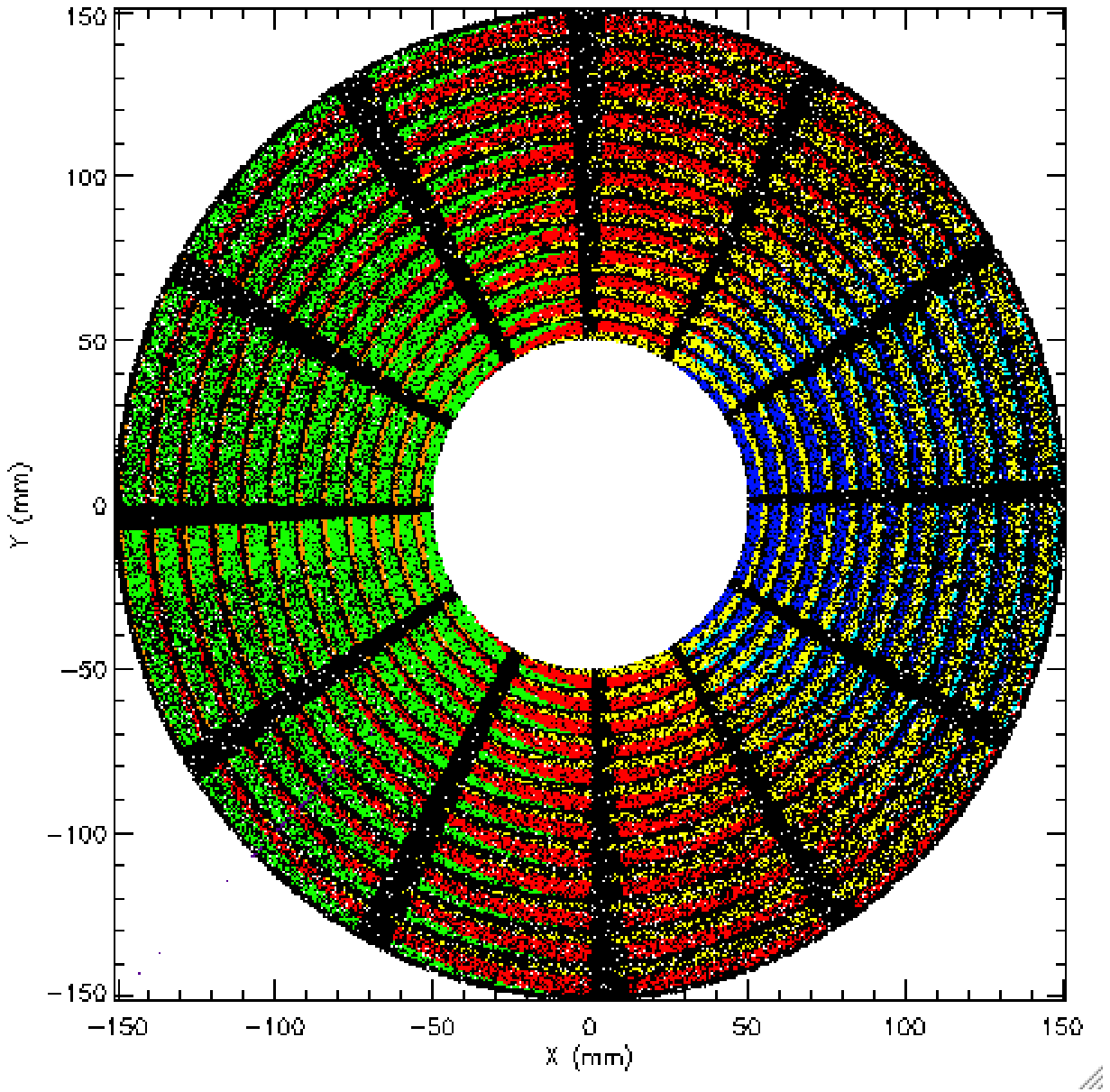}\\
		 \footnotesize A) &  \footnotesize B)\\
		 	\includegraphics[width = 0.48 \textwidth]{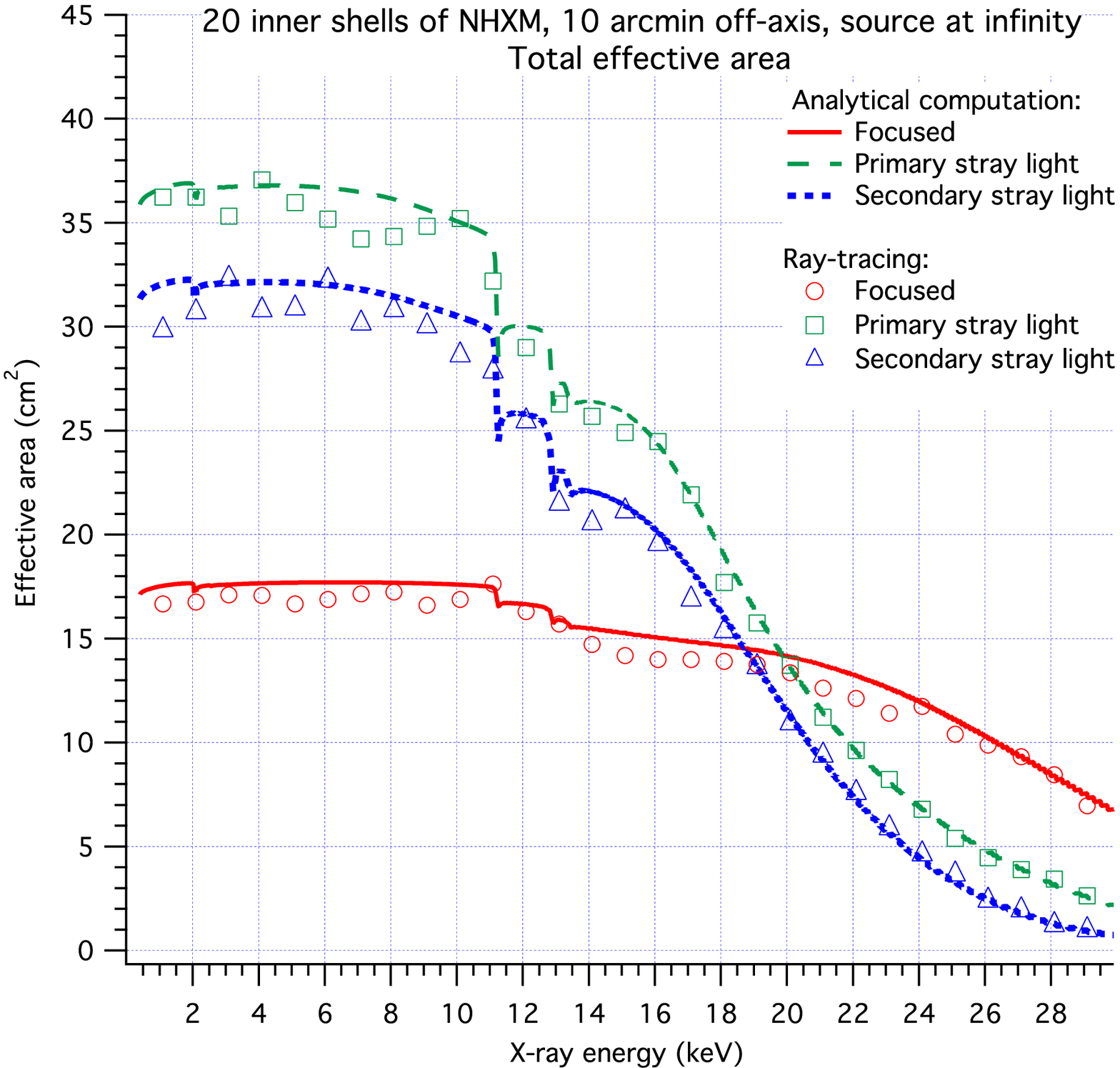}&
		 \includegraphics[width = 0.44 \textwidth]{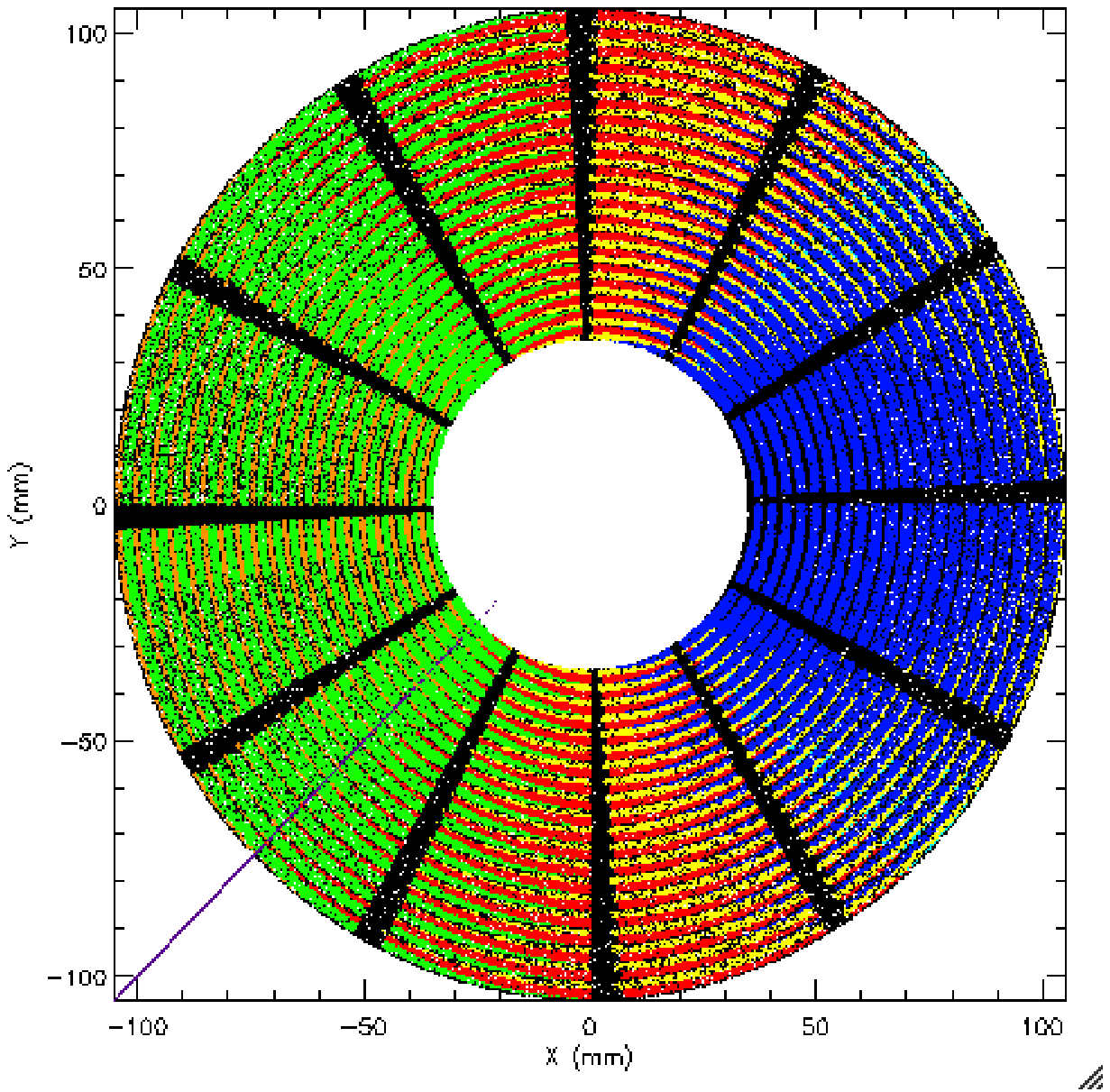}\\
		 \footnotesize C) &  \footnotesize D)
	\end{tabular}
	\caption{A) Stray light simulations for the optical module of JET-X\cite{JETX}, 25~arcmin off-axis, total effective area. The computation using the analytical method (lines) and the ray-tracing (symbols) return the same results to within a few percent.  B) Rays at the entrance pupil in colors depending on their destinations (see also Fig.~\ref{fig:entrance_pupil}). The meaning of colors is explained in the text. C) Stray light and focused effective area simulations for 20 innermost shells of the NHXM optical module\cite{NHXM}, 10~arcmin off-axis, total effective area. Also in this case the analytical method (lines) and the ray-tracing (symbols) return the same results. D) The same as B, for the optical module simulated in C.}
	\label{fig:RT_valid1}
\end{figure}

As a first example, we have computed the stray light effective area for the JET-X module\cite{JETX} ($f$= 3.5~m), 25~arcmin off-axis. The computation was referred to the total effective area, i.e., as detected over an infinitely extended focal plane. A 10\% of the aperture was assumed to be obstructed by the spider, in 12 equally-spaced spokes. The results of the analytical calculation and of the ray-tracing are shown in Fig.~\ref{fig:RT_valid1},A and are in very good accord. The analytical formulae overestimate the ray-tracing by only some percent, in agreement with our discussion in Sect.~\ref{sec:intro}. However, the application of the analytical formulae required only 15~min of computation, while more than 7 hours were needed to trace 5$\times 10^5$ rays (a number needed to reduce the statistical uncertainty) on the same IDL platform, run by a computer equipped with a commercial 2.4~GHz processor. Also, the code written to implement the analytical formulae is 8 times shorter than the code used for tracing rays.

We also show in Fig.~\ref{fig:RT_valid1},B the origin points of the rays traced. Black locations are where the rays were stopped before the first reflection or absorbed at reflection; green zones are the locations where the primary stray light was originated, while blue color marks the origin locations for secondary stray light. Primary stray light can be obstructed on the rear side of the inner shell after reflection: on the primary segment (orange) or the secondary segment (purple), or be reflected twice (red), possibly obstructed after the second reflection (light blue). The origin points of secondary stray light obstructed by the secondary segment of the inner shell are marked in yellow. The distribution of colors is in agreement with the quantitative description of vignetting given in Sect.~\ref{sec:straylight}, and also proves that all species of obstructions are accounted for in the analytical computation, thereby validating the results.

As a second example, we apply Eqs.~\ref{eq:A_2r}, \ref{eq:Asl_p_lim} and~\ref{eq:Asl_h_lim} to the NHXM hard X-ray optical module ($f$= 10~m), 10~arcmin off-axis. We have assumed a simplified coating made of 30~nm of Iridium plus 10~nm of amorphous Carbon. To furthermore simplify the visualization of the entrance pupil, we have limited the computation to the innermost 20 shells out of the 70 of the baseline design\cite{NHXM}. The result of the stray light effective area computation is shown in Fig.~\ref{fig:RT_valid1}, C. Also in this case we have derived the total effective area without limitations concerning the detector size, and the agreement with the ray tracing is still very good. In Fig.~\ref{fig:RT_valid1}, D we display the aperture pupil with colors representing the outcomes of the 5$\times 10^5$ traced rays.

\begin{figure}[hbt]
	\centering
	\begin{tabular}{ll}
		\includegraphics[width = 0.5 \textwidth]{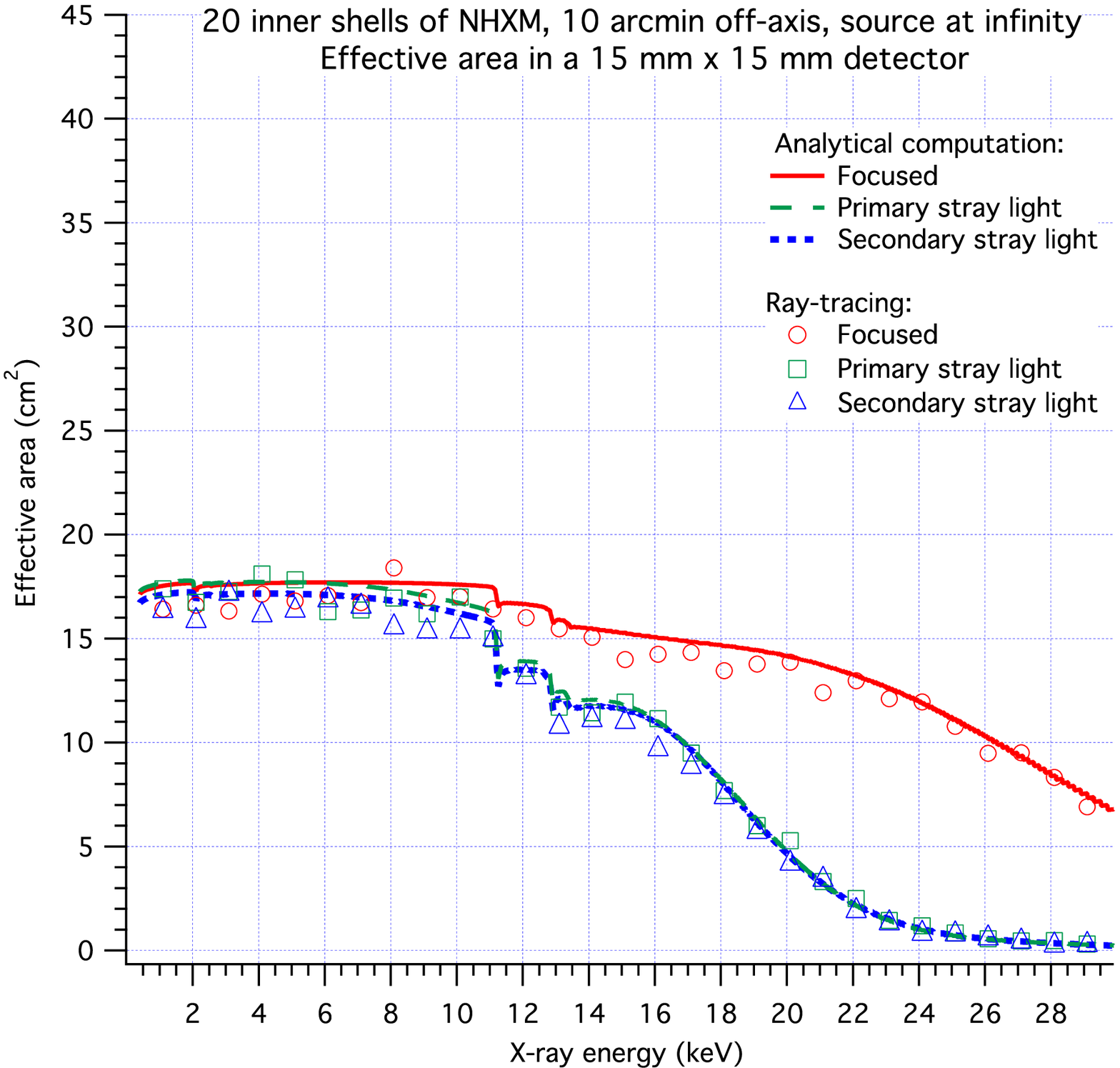}&
		 \includegraphics[width = 0.45 \textwidth]{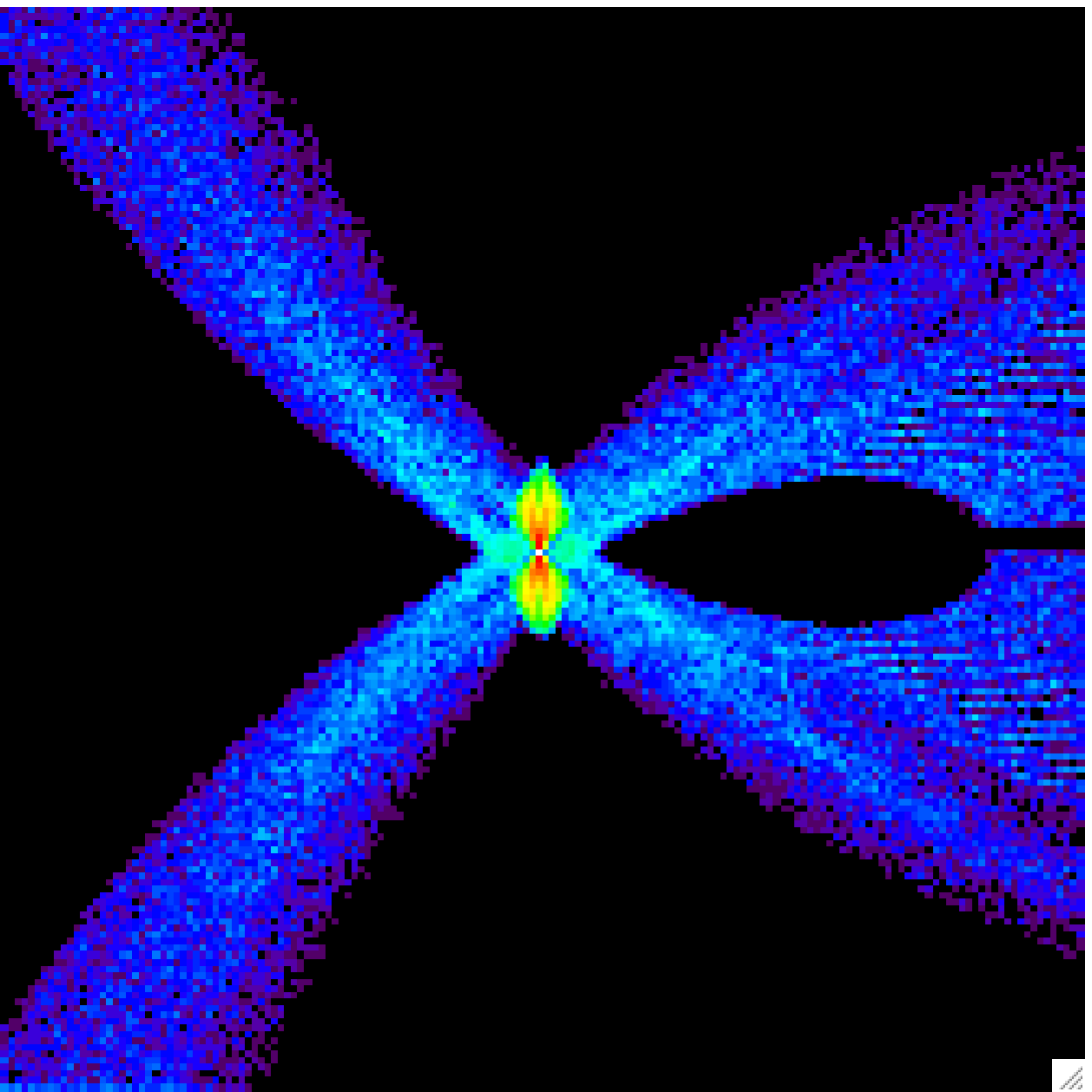}\\
		 \footnotesize A) &  \footnotesize B)
	\end{tabular}
	\caption{A) stray light and focused effective area simulations for 20 innermost shells of the NHXM optical module\cite{NHXM}, 10~arcmin off-axis. Only the stray light ending into a 15~mm $\times$ 15~mm detector area was considered. The vertical scale is the same as in Fig.~\ref{fig:RT_valid1}, C: the effective area for double reflection is unchanged, but the stray light is much lower. The ray-tracing predictions (symbols) are also in this case in good agreement with the analytical results (lines). B) The traced focal spot (center) and the stray light pattern (the big loop) in the detector area.}
	\label{fig:RT_valid2}
\end{figure}

The third and last example is shown in Fig.~\ref{fig:RT_valid2}. We have again considered the 20 innermost shells of the NHXM mirror module, but this time we have limited the computation to the effective area included in a detector of 15~mm width. The interval $[\varphi_{n,-}, \varphi_{n,+}]$ involved varies from shell to shell. The result is -- once again -- in very good agreement with the ray tracing findings. We notice that now $A^{\mathrm{SL,1}}\approx A^{\mathrm{SL,2}}$, as expected from the superposition of the primary and secondary stray light patterns from a source at infinite distance.

\section{Applications to the geometric area for stray light}\label{sec:geometric}
We hereafter deal with some applications of the formalism exposed in Sect.~\ref{sec:straylight} in the ideal case that $r_{\lambda}(\alpha)=1$ (or, at least, a constant) for all $\alpha$. To simplify the notation, we assume that $D \rightarrow \infty$, that $L_1 = L_2 = L_1^* = L_2^*$, and that we can approximate $\Phi \approx \Psi\approx \Sigma$ (Sect.~\ref{sec:vignet}): we therefore adopt $\Phi$ as unique obstruction parameter. Moreover, we do not account for limitations by the detector size, for obstruction of the supporting structures, and, since mirror modules are usually designed as obstruction-free on-axis, we assume\cite{Spiga2011} that $\alpha_0 < \Phi$.

\subsection{Geometric area, primary stray light}\label{sec:geometric1}
With the mentioned approximations, we can rewrite Eq.~\ref{eq:Asl_p} as normalized to the on-axis geometric area for double reflection, $A_{\infty}(0) = 2\pi R_0 L\alpha_0$:
\begin{equation}
	\frac{A^{\mathrm{SL,1}}_{\infty}(\theta)}{A_{\infty}(0)} = \frac{1}{\pi\alpha_0}\int_0^{\pi}[\min\left(\alpha_1, \Phi, \Phi+2\alpha_0-\alpha_1\right) -\max\left(2\alpha_0-\alpha_1, \alpha_1-\Phi, 0\right)]_{\ge 0} \,\mbox{d}\varphi.
	\label{eq:Asl_geo}
\end{equation} 
In order to explicitly solve Eq.~\ref{eq:Asl_geo}, we distinguish between two main cases in the following paragraphs.

\subsubsection{SL1, tight nesting: $\Phi/2 < \alpha_0$}\label{sec:SL1_tight}
In addition to $\alpha_0 < \Phi$, we initially assume that $\Phi/2 < \alpha_0$, which implies that 
\begin{equation}
	0<\Phi-\alpha_0<\Phi/2<\alpha_0<\Phi. 
	\label{eq:chain1}
\end{equation}
We can therefore distinguish 5 different angular regimes:
\begin{itemize}
\item{$0< \theta < \Phi-\alpha_0$: in this angular range, $\alpha_1 <\Phi$ everywhere. Since $\theta < \Phi/2$, we also have that $\alpha_1 <\Phi+2\alpha_0-\alpha_1$ and $2\alpha_0-\alpha_1 > \alpha_1-\Phi$; hence, there is no obstruction, so Eq.~\ref{eq:Asl_geo} takes the simple form:
\begin{equation}
	\frac{A^{\mathrm{SL,1}}_{\infty}(\theta)}{A_{\infty}(0)} =  \frac{2}{\pi\alpha_0}\int_{\pi/2}^{\pi}(\alpha_1-\alpha_0)\,\mbox{d}\varphi = \frac{2\theta}{\pi\alpha_0},
	\label{eq:Asl_geo1}
\end{equation}
where the lower integration limit is modified to keep the integrand non-negative. In this range of off-axis angles, Eq.~\ref{eq:Asl_geo}  correctly returns the complement to the unity of the normalized, geometric, unobstructed area for double reflection\cite{VanSpey, Spiga2009}.}
\item{$\Phi-\alpha_0< \theta < \Phi/2$: this time we have $\Phi<\alpha_1$, i.e., obstruction of the second species for $\cos\varphi < - (\Phi-\alpha_0)/\theta$, and Eq.~\ref{eq:Asl_geo} becomes
\begin{equation}
		\frac{A^{\mathrm{SL,1}}_{\infty}(\theta)}{A_{\infty}(0)} =   \frac{2}{\pi\alpha_0}\left[\int_{\frac{\pi}{2}}^{\pi-\arccos[(\Phi-\alpha_0)/\theta]}\!\!\hspace{-1.8cm}(\alpha_1-\alpha_0)\,\mbox{d}\varphi + \frac{1}{2}\int^{\pi}_{\pi-\arccos[(\Phi-\alpha_0)/\theta]} \hspace{-1.8cm}(\Phi-2\alpha_0+\alpha_1)\,\mbox{d}\varphi\right],
	\label{eq:Asl_geo3}
\end{equation}
where the integrands are always non-negative because $\cos\varphi < 0$. Equation~\ref{eq:Asl_geo3} can be solved as
\begin{equation}
		\frac{A^{\mathrm{SL,1}}_{\infty}(\theta)}{A_{\infty}(0)} =  \frac{2\theta}{\pi\alpha_0}\left[1-\frac{1}{2} S\left(\frac{\Phi-\alpha_0}{\theta}\right)\right],
	\label{eq:Asl_geo3bis}
\end{equation}
where we have introduced the non-negative $S$ function (already defined in \cite{Spiga2011})
\begin{equation}
		S(x) = \sqrt{1-x^2}-x\arccos x , \hspace{1cm} \mbox{with $0\le x \le 1$}.
	\label{eq:S_funct}
\end{equation}}
\item{$\Phi/2 < \theta <\alpha_0$: we still have $\Phi < \Phi+2\alpha_0-\alpha_1$ everywhere. However, for $\cos\varphi < -\Phi/2\theta$ we also have $\alpha_1-\Phi > 2\alpha_0-\alpha_1$ and we begin experiencing obstruction of the first kind. Equation~\ref{eq:Asl_geo} now turns into
\begin{equation}
		\frac{A^{\mathrm{SL,1}}_{\infty}(\theta)}{A_{\infty}(0)} =   \frac{2}{\pi\alpha_0}\left[\int_{\frac{\pi}{2}}^{\pi-\arccos[(\Phi-\alpha_0)/\theta]}\!\!\hspace{-1.8cm}(\alpha_1-\alpha_0)\,\mbox{d}\varphi + \frac{1}{2}\int^{\pi-\arccos(\Phi/2\theta)}_{\pi-\arccos[(\Phi-\alpha_0)/\theta]} \hspace{-1.8cm}(\Phi-2\alpha_0+\alpha_1)\,\mbox{d}\varphi+ \frac{1}{2}\int^{\pi}_{\pi-\arccos(\Phi/2\theta)} \hspace{-1.5cm}(2\Phi-\alpha_1)\,\mbox{d}\varphi\right],
	\label{eq:Asl_geo4int}
\end{equation} 
and, after a few passages, the expression can be written in an explicit form: 
\begin{equation}
		\frac{A^{\mathrm{SL,1}}_{\infty}(\theta)}{A_{\infty}(0)} =  \frac{2\theta}{\pi\alpha_0}\left[1-\frac{1}{2} S\left(\frac{\Phi-\alpha_0}{\theta}\right)-S\left(\frac{\Phi}{2\theta}\right)\right].
	\label{eq:Asl_geo4}
\end{equation}}
\item{$\alpha_0<\theta <\Phi$: for $\cos\varphi < -\alpha_0/\theta$ the first term of the integrand in Eq.~\ref{eq:Asl_geo} becomes $\Phi+2\alpha_0-\alpha_1$. This denotes the appearance of the obstruction of the third kind: the final expression for these values of $\theta$ is
\begin{equation}
		\frac{A^{\mathrm{SL,1}}_{\infty}(\theta)}{A_{\infty}(0)} =  \frac{2\theta}{\pi\alpha_0}\left[1-\frac{1}{2} S\left(\frac{\Phi-\alpha_0}{\theta}\right)-S\left(\frac{\Phi}{2\theta}\right)-\frac{1}{2}S\left(\frac{\alpha_0}{\theta}\right)\right].
	\label{eq:Asl_geo5}
\end{equation}}
\item{$\theta >\Phi$: the expression of the integrand remains unchanged, but we have to modify the upper integration limit to avoid negative contributions to the integral, and the final result is:
\begin{equation}
		\frac{A^{\mathrm{SL,1}}_{\infty}(\theta)}{A_{\infty}(0)} =  \frac{2\theta}{\pi\alpha_0}\left[1-\frac{1}{2} S\left(\frac{\Phi-\alpha_0}{\theta}\right)-S\left(\frac{\Phi}{2\theta}\right)-\frac{1}{2}S\left(\frac{\alpha_0}{\theta}\right)+S\left(\frac{\Phi}{\theta}\right)\right].
	\label{eq:Asl_geo6}
\end{equation}}
\end{itemize} 

\begin{figure}[th]
	\centering
		\includegraphics[width = 0.8 \textwidth]{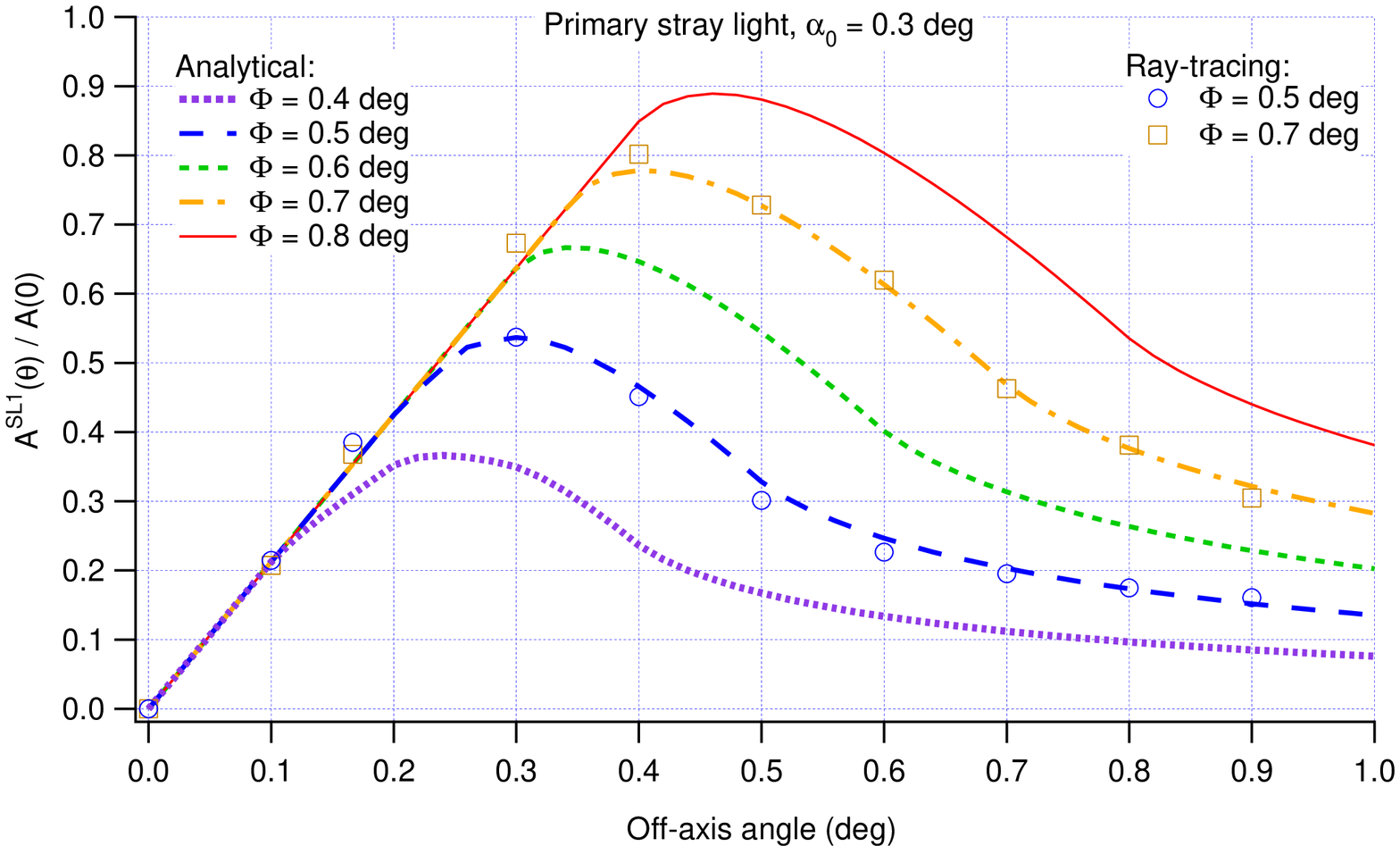}
	\caption{Total, normalized geometric area for {\it primary} stray light of a mirror shell with $\alpha_0$ = 0.3~deg, as a function of the off-axis angle, for variable values of the obstruction parameter $\Phi$ (assuming that $\Phi \approx \Psi \approx \Sigma$). The expressions in Sect.~\ref{sec:geometric1} (lines) have been used in the respective intervals of validity. Some values computed by ray-tracing (symbols) are also shown for verification.}
	\label{fig:geom_area1}
\end{figure} 

\subsubsection{SL1, loose nesting: $\alpha_0 < \Phi/2$}\label{sec:SL1_loose}
We now admit that $\alpha_0 < \Phi/2$, which implies $\Phi/2 < \Phi-\alpha_0 <\Phi$. We have the following results in different ranges of the off-axis angle:
\begin{itemize}
\item{$0<\theta <\Phi/2$: as this implies $\theta < \Phi/2 < \Phi-\alpha_0$, we have $\alpha_1 < \Phi$ everywhere. Moreover, $2\alpha_0 +\Phi> 2\alpha_1$, and then the result is equal to Eq.~\ref{eq:Asl_geo1}.}
\item{$\Phi/2<\theta <\Phi-\alpha_0$: in this case we still have $\alpha_1 < \Phi$ everywhere, but there is some interval of $\varphi<\sim \pi$ where $\Phi< -2\theta\cos\varphi$; therefore, in this range of $\varphi$ we start to have some stray light obstruction of the 1-st and the 3-rd kind. Equation~\ref{eq:Asl_geo} becomes
\begin{equation}
		\frac{A^{\mathrm{SL,1}}_{\infty}(\theta)}{A_{\infty}(0)} =  \frac{2}{\pi\alpha_0}\left[\int_{\frac{\pi}{2}}^{\pi-\arccos(\Phi/2\theta)}\!\!\hspace{-1.6cm}(\alpha_1-\alpha_0)\,\mbox{d}\varphi + \int^{\pi}_{\pi-\arccos(\Phi/2\theta)}\hspace{-1.6cm}(\Phi+\alpha_0-\alpha_1)\,\mbox{d}\varphi\right],
	\label{eq:Asl_geo2}
\end{equation}
and since $\theta <\Phi-\alpha_0<\Phi$, we second integrand is always non-negative. Using the expression of the $S$ function (Eq.~\ref{eq:S_funct}), the last expression can be rewritten as 
\begin{equation}
		\frac{A^{\mathrm{SL,1}}_{\infty}(\theta)}{A_{\infty}(0)} =  \frac{2\theta}{\pi\alpha_0}\left[1-2 S\left(\frac{\Phi}{2\theta}\right)\right].
	\label{eq:Asl_geo2bis}
\end{equation}}
\item{$\Phi-\alpha_0<\theta<\Phi $: in this case we potentially have an interval of polar angles affected by vignetting of the second kind. Indeed, this occurs only if $\Phi < \Phi+2\alpha_0-\alpha_1$, i.e., only if $\cos\varphi > -\alpha_0/\theta$. On the other hand, this would also require $\Phi < \alpha_1$, which is equivalent to $\cos\varphi < -(\Phi-\alpha_0)/\theta$. Fulfilling the two conditions at the same time would require $ -\alpha_0 <  -(\Phi-\alpha_0)$, which is impossible since we initially supposed that $\alpha_0<\Phi/2$. We conclude that in this interval of off-axis angles the geometric area is still described by Eq.~\ref{eq:Asl_geo2bis}.
\item{$\theta>\Phi$: while the form of the integrand remains the same of Eq.~\ref{eq:Asl_geo2}, the second integrand has to be set to zero for $\cos\varphi <-\Phi/\theta$ to avoid negative values, i.e., 
\begin{equation}
		\frac{A^{\mathrm{SL,1}}_{\infty}(\theta)}{A_{\infty}(0)} =  \frac{2}{\pi\alpha_0}\left[\int_{\frac{\pi}{2}}^{\pi-\arccos(\Phi/2\theta)}\!\!\hspace{-1.6cm}(-\theta\cos\varphi)\,\mbox{d}\varphi + \int_{\pi-\arccos(\Phi/2\theta)}^{\pi-\arccos(\Phi/\theta)}\hspace{-1.6cm}(\Phi+\theta\cos\varphi)\,\mbox{d}\varphi\right],
	\label{eq:Asl_geo7}
\end{equation}
which can be easily written as 
\begin{equation}
		\frac{A^{\mathrm{SL,1}}_{\infty}(\theta)}{A_{\infty}(0)} = \frac{2\theta}{\pi\alpha_0}\left[1-2 S\left(\frac{\Phi}{2\theta}\right)+S\left(\frac{\Phi}{\theta}\right)\right].
	\label{eq:Asl_geo7bis}
\end{equation}}
}\end{itemize}

We notice that Eq.~\ref{eq:Asl_geo3bis} and~\ref{eq:Asl_geo4} are no longer valid when $\alpha_0 = \Phi/2$, since their domain collapse to the single point $\theta = \Phi/2$. Moreover, when $\alpha_0 = \Phi/2$, Eqs.~\ref{eq:Asl_geo5} and~\ref{eq:Asl_geo6} correctly merge with Eqs.~\ref{eq:Asl_geo2bis} and~\ref{eq:Asl_geo7bis}, respectively. 

We show in Fig.~\ref{fig:geom_area1} some examples of normalized geometric area curves, computed using Eqs.~\ref{eq:Asl_geo1} to~\ref{eq:Asl_geo7bis} in the respective $\theta$ domains. The geometric area profiles exhibit a characteristic peak followed by a gradual decrease for increasing off-axis angles, in qualitative agreement with ray-traced data reported for the case of SIMBOL-X telescope\cite{Cusumano2007}. Comparison with some ray-tracing results (symbols in Fig.~\ref{fig:geom_area1}) shows an excellent agreement.

\subsection{Geometric area, secondary stray light} \label{sec:geometric2}
Within the same approximations, Eq.~\ref{eq:Asl_h} yields for the normalized geometric area for secondary stray light, once normalized to the on-axis geometric area for double reflection,
\begin{equation}
	\frac{A^{\mathrm{SL,2}}_{\infty}(\theta)}{A_{\infty}(0)} = \frac{1}{\pi\alpha_0}\int_0^{\pi}[\min\left(2\alpha_0+\alpha_1, \Phi, \Phi-\alpha_1\right) -\max\left(-\alpha_1, 2\alpha_0-\Phi+\alpha_1, 0\right)]_{\ge 0} \,\mbox{d}\varphi,
	\label{eq:Asl2_geo}
\end{equation} 
where we used the relations $\beta_2 = 2\alpha_0+\alpha_1$ and $\beta_1 = -\alpha_1$. We once more have to discriminate between the cases that $\Phi/2 < \alpha_0$ or $\Phi/2 > \alpha_0$.

\subsubsection{SL2, tight nesting: $\Phi/2 < \alpha_0$}\label{sec:SL2_tight}
We firstly consider the case of $\Phi/2 < \alpha_0$, which implies the inequalities:
\begin{equation}
	0 < 2\alpha_0-\Phi < \alpha_0 < 2\alpha_0-\Phi/2< 3\alpha_0-\Phi <2\alpha_0.
	\label{eq:chain2}
\end{equation}
Also Eq.~\ref{eq:Asl2_geo} takes different expressions in different intervals of $\theta$.

\begin{itemize}
		\item{$0 < \theta <2\alpha_0-\Phi$: in this range of off-axis angles we have $2\alpha_0+\alpha_1 > \Phi > \Phi-\alpha_1$, and $2\alpha_0-\Phi+\alpha_1 > 0$ for all $\varphi$. Equation~\ref{eq:Asl2_geo} then becomes
\begin{equation}
	\frac{A^{\mathrm{SL,2}}_{\infty}(\theta)}{A_{\infty}(0)} = \frac{2}{\pi\alpha_0}\int_0^{\pi} [\Phi-2\alpha_0+\theta\cos\varphi]_{\ge 0}\,\mbox{d}\varphi  = 0,
	\label{eq:Asl2_geo1}
\end{equation}		
because the expression in $[\,]_{\ge 0}$ is always negative; hence, the integrand has been zeroed. As expected, the secondary stray light is completely obstructed at the exit pupil for small off-axis values.}
	\item{$2\alpha_0-\Phi < \theta <\alpha_0$: the expression of the integrand is the same as in Eq.~\ref{eq:Asl2_geo1}, but this time it becomes positive for $\cos\varphi >(2\alpha_0-\Phi)/\theta$, and we obtain
\begin{equation}
	\frac{A^{\mathrm{SL,2}}_{\infty}(\theta)}{A_{\infty}(0)} = \frac{2}{\pi\alpha_0}\int_0^{\arccos[(2\alpha_0-\Phi)/\theta]}\hspace{-1.8cm} (\Phi-2\alpha_0+\theta\cos\varphi) \,\mbox{d}\varphi.
	\label{eq:Asl2_geo2}
\end{equation}	
Developing the computation we obtain the result in terms of the $S$ function (Eq.~\ref{eq:S_funct}):
	\begin{equation}
	\frac{A^{\mathrm{SL,2}}_{\infty}(\theta)}{A_{\infty}(0)} = \frac{2\theta}{\pi\alpha_0}S\left(\frac{2\alpha_0-\Phi}{\theta}\right).
	\label{eq:Asl2_geo2bis}
\end{equation}}
\item{$\alpha_0< \theta <2\alpha_0-\Phi/2$: in this case, we have $\Phi < \Phi-\alpha_1$ for $\cos\varphi >\alpha_0/\theta$, where the vignetting of the second kind takes over the first kind. However, $-\alpha_1 < 2\alpha_0-\Phi+\alpha_1$ everywhere, and we do not need to change the second term in Eq.~\ref{eq:Asl2_geo}. The normalized area can now be written as
\begin{equation}
	\frac{A^{\mathrm{SL,2}}_{\infty}(\theta)}{A_{\infty}(0)} = \frac{2}{\pi\alpha_0}\left[\frac{1}{2}\int_0^{\arccos(\alpha_0/\theta)}\hspace{-1.4cm}(2\Phi-3\alpha_0+\theta\cos\varphi)\,\mbox{d}\varphi +\int_{\arccos(\alpha_0/\theta)}^{\arccos[(2\alpha_0-\Phi)/\theta]}\hspace{-1.8cm} (\Phi-2\alpha_0+\theta\cos\varphi) \,\mbox{d}\varphi\right],
	\label{eq:Asl2_geo3}
\end{equation}	
and the first integrand is non-negative in the integration set. Hence, solving the integrals we get
\begin{equation}
	\frac{A^{\mathrm{SL,2}}_{\infty}(\theta)}{A_{\infty}(0)} = \frac{2\theta}{\pi\alpha_0}\left[S\left(\frac{2\alpha_0-\Phi}{\theta}\right)-\frac{1}{2}S\left(\frac{\alpha_0}{\theta}\right) \right].
	\label{eq:Asl2_geo3bis}
\end{equation}}
\item{$2\alpha_0-\Phi/2< \theta < 3\Phi-\alpha_0$: there is now an interval of polar angles, $\cos\varphi > (2\alpha_0-\Phi/2)/\theta$, affected by obstruction on the rear side of the reflective mirror shell ($-\alpha_1 > 2\alpha_0-\Phi+\alpha_1$). In the same region, the obstruction of the second kind is still effective since $\Phi < 2\alpha_0+\alpha_1 < \Phi-\alpha_1$: developing the passages, we remain with
\begin{equation}
	\frac{A^{\mathrm{SL,2}}_{\infty}(\theta)}{A_{\infty}(0)} = \frac{2\theta}{\pi\alpha_0}\left[S\left(\frac{2\alpha_0-\Phi}{\theta}\right)-\frac{1}{2}S\left(\frac{\alpha_0}{\theta}\right) -S\left(\frac{2\alpha_0-\Phi/2}{\theta}\right)\right].
	\label{eq:Asl2_geo4}
\end{equation}}
\begin{figure}[thb]
	\centering
	\includegraphics[width = 0.8 \textwidth]{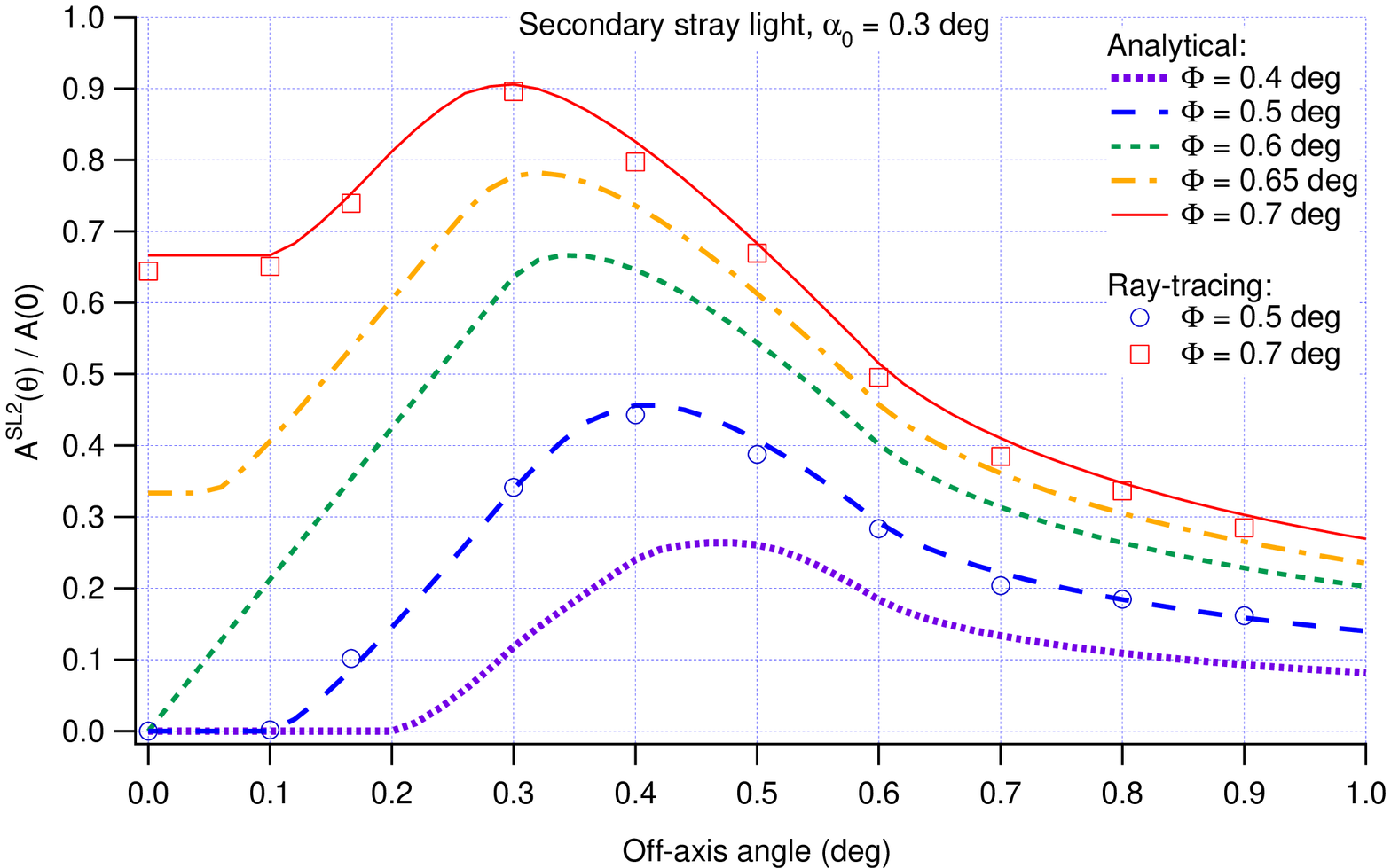}
	\caption{Total, normalized geometric area for {\it secondary} stray light of a mirror shell with $\alpha_0$ = 0.3~deg, as a function of the off-axis angle, for variable values of the obstruction parameter $\Phi$ (assuming that $\Phi \approx \Psi \approx \Sigma$). The expressions in Sect.~\ref{sec:geometric2} (lines) have been used in the respective intervals of validity. Some values computed by ray-tracing (symbols) are also shown for verification.}
	\label{fig:geom_area2}
\end{figure} 
\item{$ 3\alpha_0-\Phi < \theta < 2\alpha_0$: in the polar angle range defined by $\cos\varphi > (3\alpha_0-\Phi)/\theta$ the obstruction is, this time, no longer effective because $2\alpha_0+\alpha_1 < \Phi$: in this range of $\theta$ values, the expression to be adopted turns out to be 
\begin{equation}
	\frac{A^{\mathrm{SL,2}}_{\infty}(\theta)}{A_{\infty}(0)} = \frac{2\theta}{\pi\alpha_0}\left[S\left(\frac{2\alpha_0-\Phi}{\theta}\right)-\frac{1}{2}S\left(\frac{\alpha_0}{\theta}\right) -S\left(\frac{2\alpha_0-\Phi/2}{\theta}\right)-\frac{1}{2}S\left(\frac{3\alpha_0-\Phi}{\theta}\right)\right].
	\label{eq:Asl2_geo5}
\end{equation}}
\item{$\theta > 2\alpha_0$: the integrand expression is exactly the same as in the previous case, but we just have to change the lower integration limit from 0 to $\arccos(2\alpha_0/\theta)$ to avoid negative contribution to the effective area. So we obtain the result, by adding a last term:
\begin{equation}
	\frac{A^{\mathrm{SL,2}}_{\infty}(\theta)}{A_{\infty}(0)} = \frac{2\theta}{\pi\alpha_0}\left[S\left(\frac{2\alpha_0-\Phi}{\theta}\right)-\frac{1}{2}S\left(\frac{\alpha_0}{\theta}\right) -S\left(\frac{2\alpha_0-\Phi/2}{\theta}\right)-\frac{1}{2}S\left(\frac{3\alpha_0-\Phi}{\theta}\right)+S\left(\frac{2\alpha_0}{\theta}\right)\right].
	\label{eq:Asl2_geo6}
\end{equation}}
\end{itemize}

\subsubsection{SL2, loose nesting: $\alpha_0 < \Phi/2$}\label{sec:SL2_loose}
Finally, we take into account the case of $2\alpha_0 < \Phi$. However, to avoid complicating further the expressions, we additionally assume an upper bound $\Phi < 5\alpha_0/2$. This implies the ordering 
\begin{equation}
	0< \Phi-2\alpha_0<3\alpha_0-\Phi < 2\alpha_0-\Phi/2 < \alpha_0< 2\alpha_0,
	\label{eq:chain3}
\end{equation}
and, repeating the reasoning we did for the previous cases, we obtain the following expressions:
	\begin{itemize}
		\item{$0<\theta< \Phi-2\alpha_0$: for this case, the form of the integrand is identical to the one obtained in Eq.~\ref{eq:Asl2_geo1}, with the difference that this time $\Phi-2\alpha_0 >0$ and so the integrand is always non-negative. The result is
\begin{equation}
	\frac{A^{\mathrm{SL,2}}_{\infty}(\theta)}{A_{\infty}(0)} = \frac{2(\Phi-2\alpha_0)}{\alpha_0}.
	\label{eq:Asl2_geo7}
\end{equation}}
\item{$\Phi-2\alpha_0 < \theta <3\alpha_0-\Phi$: the integrand has the same form as before, but in order to ensure that it is always non-negative we have to change the upper integration limit as we did for Eq.~\ref{eq:Asl2_geo2}, and the result is
\begin{equation}
	\frac{A^{\mathrm{SL,2}}_{\infty}(\theta)}{A_{\infty}(0)} = \frac{2(\Phi-2\alpha_0)}{\alpha_0}+ \frac{2\theta}{\pi\alpha_0}S\left(\frac{\Phi-2\alpha_0}{\theta}\right).
	\label{eq:Asl2_geo8}
\end{equation}}
\item{$3\alpha_0-\Phi < \theta <2\alpha_0-\Phi/2$: the first term in the integrand of Eq.~\ref{eq:Asl2_geo} is still $\Phi-\alpha_1$. However, for  $\cos\varphi > (3\alpha_0- \Phi)/\theta$ the second term becomes negative and should be replaced by zero. Hence, the expression valid in this range is
\begin{equation}
	\frac{A^{\mathrm{SL,2}}_{\infty}(\theta)}{A_{\infty}(0)} = \frac{2(\Phi-2\alpha_0)}{\alpha_0}+ \frac{2\theta}{\pi\alpha_0}\left[S\left(\frac{\Phi-2\alpha_0}{\theta}\right)-\frac{1}{2}S\left(\frac{3\alpha_0-\Phi}{\theta}\right)\right].
	\label{eq:Asl2_geo9}
\end{equation}}
\item{$2\alpha_0-\Phi/2 < \theta <\alpha_0$: for $\cos\varphi > (2\alpha_0-\Phi/2)/\theta$ the first term in the integrand of Eq.~\ref{eq:Asl2_geo} is now $2\alpha_0+\alpha_1$, while the second term is still zero. The result valid in this range of $\theta$ values is
\begin{equation}
	\frac{A^{\mathrm{SL,2}}_{\infty}(\theta)}{A_{\infty}(0)} = \frac{2(\Phi-2\alpha_0)}{\alpha_0}+ \frac{2\theta}{\pi\alpha_0}\left[S\left(\frac{\Phi-2\alpha_0}{\theta}\right)-\frac{1}{2}S\left(\frac{3\alpha_0-\Phi}{\theta}\right)-S\left(\frac{2\alpha_0-\Phi/2}{\theta}\right)\right].
	\label{eq:Asl2_geo10}
\end{equation}}
\item{$\alpha_0 < \theta <2\alpha_0$: for $\cos\varphi > \alpha_0/\theta$ the first term in Eq.~\ref{eq:Asl2_geo} remains $2\alpha_0+\alpha_1$ and the second one is now $-\alpha_1$. Developing the computation, the result is
\begin{equation}
	\frac{A^{\mathrm{SL,2}}_{\infty}(\theta)}{A_{\infty}(0)} = \frac{2(\Phi-2\alpha_0)}{\alpha_0}+ \frac{2\theta}{\pi\alpha_0}\left[S\left(\frac{\Phi-2\alpha_0}{\theta}\right)-\frac{1}{2}S\left(\frac{3\alpha_0-\Phi}{\theta}\right)-S\left(\frac{2\alpha_0-\Phi/2}{\theta}\right)-\frac{1}{2}S\left(\frac{\alpha_0}{\theta}\right)\right].
	\label{eq:Asl2_geo11}
\end{equation}}
\item{$\theta > 2\alpha_0$: with respect to the previous case, we just have to modify the upper integration limit to ensure that the integrand, $2\alpha_0+2\alpha_1$, remains non-negative. The final result is obtained by adding $S(2\alpha_0/\theta)$ to the sum in $[\,]$ in Eq.~\ref{eq:Asl2_geo11}.}
	\end{itemize}

As for primary stray light, the results found in Sect.~\ref{sec:SL2_tight} and~\ref{sec:SL2_loose} merge in the limit case that $\alpha_0 = \Phi/2$. If is interesting to note that for this value of $\Phi$ the results also merge with the expressions for primary stray light (Sect.~\ref{sec:SL1_tight} and~\ref{sec:SL1_loose}). We show in Fig.~\ref{fig:geom_area2} some examples of normalized geometric area curves, computed using Eqs.~\ref{eq:Asl2_geo1} to~\ref{eq:Asl2_geo11} in the respective $\theta$ domains. Comparison with some ray-tracing results (symbols in Fig.~\ref{fig:geom_area2}) shows an excellent agreement.

\section{Conclusions and mirror module design considerations}\label{sec:conclusions}
In this paper we have reviewed the possible sources of obstruction for focused and stray rays in nested modules of Wolter-I mirrors. We have thereby found integral formulae to compute, in addition to the already known expression for the double-reflection (focused) intensity, the effective area for stray light off the primary (Eq.~\ref{eq:Asl_p}) and the secondary (Eq.~\ref{eq:Asl_h}) mirror segments, also accounting for the finite size of the detector (Eqs.~\ref{eq:Asl_p_lim} and~\ref{eq:Asl_h_lim}). The predictions are in very good agreement with the ray-tracing findings. In the ideal case of a mirror with constant reflectivity, algebraic expressions for the geometric area could be provided in different ranges of off-axis angles.

The formalism provided here can be useful in designing a mirror module maximizing the focused effective area and, at the same time, minimizing the stray light impact. In fact, the solution of designing a completely obstruction-free mirror module within the field of view\cite{Spiga2011} might leave too much spacing for the stray light to propagate throughout the mirror nesting. In contrast, the formalism provided here enables, from a given mirror module design, not only a fast assessment of the stray light impact from off-axis sources; it also returned useful relations between the tolerable stray-light magnitude and the obstruction parameters. They therefore provide a way to establish the optimal obstruction to minimize the effective area for stray light while preserving the required effective area in the field of view. Should these formulae be solved numerically for $\Phi$, $\Psi$, and $\Sigma$, the complex task of mirror module design problem could be solved easily without the need to run a complex ray-tracing program. Finally, the same method might be applied to the problem of designing an X-ray baffle, only by a simple re-definition of the obstruction parameters.

As a future development of this work, the formalism might be extended to include the case of segmented mirrors, as the ones foreseen for the ATHENA telescope. However, this kind of optics usually include stiffening ribs that represent a further source of obstruction, and the problem becomes more complicated to treat analytically.

\bibliographystyle{spiebib}

\end{document}